\documentclass[prb,superscriptaddress,showpacs,floatfix,twocolumn]{revtex4}
\usepackage{amssymb}
\usepackage{amsmath}
\usepackage{graphicx}

\begin{document}

\title{Collective modes of $CP^{3}$ Skyrmion crystals in quantum Hall
ferromagnets }
\author{R. C\^{o}t\'{e}}
\affiliation{D\'{e}partement de physique, Universit\'{e} de Sherbrooke, Sherbrooke, Qu%
\'{e}bec, Canada, J1K 2R1}
\author{D. B. Boisvert}
\affiliation{D\'{e}partement de physique, Universit\'{e} de Sherbrooke, Sherbrooke, Qu%
\'{e}bec, Canada, J1K 2R1}
\author{J. Bourassa}
\affiliation{D\'{e}partement de physique, Universit\'{e} de Sherbrooke, Sherbrooke, Qu%
\'{e}bec, Canada, J1K 2R1}
\author{M. Boissonneault}
\affiliation{D\'{e}partement de physique, Universit\'{e} de Sherbrooke, Sherbrooke, Qu%
\'{e}bec, Canada, J1K 2R1}
\email{Rene.Cote@USherbrooke.ca}
\author{H. A. Fertig}
\affiliation{Department of Physics, Indiana University, Bloomington, Indiana 47405 and \\
Department of Physics, Technion, Haifa 32000 Israel}

\begin{abstract}
The two-dimensional electron gas (2DEG) in a bilayer quantum Hall system can
sustain an interlayer coherence at filling factor $\nu =1$ even in the
absence of tunneling between the layers. This system, which can be described
as a quantum Hall pseudospin ferromagnet, has low-energy charged excitations
which may carry textures in real spin or pseudospin. Away from filling
factor $\nu =1,$ a finite density of these is present in the ground state of
the 2DEG and forms a crystal. Depending on the relative size of the various
energy scales, such as tunneling ($\Delta _{SAS}$), Zeeman coupling ($\Delta
_{Z}$) or electrical bias ($\Delta _{b}$), these textured crystal states can
involve spin, pseudospin, or both intertwined. This last case is a
\textquotedblleft CP$^{3}$ Skyrmion Crystal\textquotedblright .\ In this
article, we present a comprehensive numerical study of the collective
excitations of these textured crystals using the Generalized Random-Phase
Approximation. For the pure spin case, at finite Zeeman coupling the state
is a Skyrmion crystal with a gapless phonon mode, and a separate Goldstone
mode that arises from a broken $U\left( 1\right) $ symmetry. At zero Zeeman
coupling, we demonstrate that the constituent Skyrmions break up, and the
resulting state is a meron crystal with \textit{four} gapless modes. In
contrast, a pure pseudospin Skyrme crystal at finite tunneling has only the
phonon mode. For $\Delta _{SAS}\rightarrow 0,$ the state evolves into a
meron crystal and supports an extra gapless ($U(1)$) mode in addition to the
phonon. For a $CP^{3}$ Skyrmion crystal, we find a $U(1)$ gapless mode in
the presence of non-vanishing symmetry-breaking fields $\Delta _{SAS}$, $%
\Delta _{Z},$ and $\Delta _{b}$. In addition, a second mode with a very
small gap is present in the spectrum. We present dispersion relations for
the different low-energy modes of these various crystals as well as their
physical interpretations.
\end{abstract}

\date{\today }
\maketitle

\section{Introduction}

The two-dimensional electron gas (2DEG) in a bilayer quantum Hall system has
a broken-symmetry ground state at filling factor $\nu =1$ that can be
described as a pseudospin ferromagnet. This state is characterized by finite
interlayer coherence even in the absence of tunneling. Interlayer coherence
is maintained if the interlayer separation $d$ is lower than a critical
separation which is of the order of $d_{c}\approx 1.25\ell $ at zero
electrical bias where $\ell =\sqrt{\hslash c/eB}$ is the magnetic length.
This state has been extensively studied, both experimentally and
theoretically (for a review, see Ref. \onlinecite{review}). Its collective
excitation\cite{fertigdispersion}, which has been detected experimentally%
\cite{spielmanpseudospinmode}, is a pseudospin wave mode in which the
pseudospins precess around their equilibrium orientations.

In most studies of the interlayer coherent states at $\nu =1$ or at other
filling factors where states such as Wigner crystals, Skyrmion crystals or
bilayer stripes are predicted, it is assumed that the 2DEG is spin polarized
so that the non-interacting 2DEG can be mapped onto a two-level system with
bonding and anti-bonding states. Recent experiments\cite%
{kumadaprl,spielmanprl}, however, cast some doubt on the validity of this
assumption. To explain these recent experimental findings, it seems that the
ground state should be partially spin depolarized.

A previous study\cite{cotecp3} (which we refer to as Paper I\ in the rest of
this article) presented various zero-temperature phases of the 2DEG in a
bilayer quantum Hall system when the filling factor is varied around $\nu =1$%
, so that a finite density of charged excitations is introduced in the
ground state. Working in the Hartree-Fock approximation, various symmetry
breaking fields were present in the system: tunnelling $\left( \Delta
_{SAS}\right) $, Zeeman $\left( \Delta _{Z}\right) $ and electrical bias $%
\left( \Delta _{b}\right) $, over a range of experimentally relevant filling
factors ($0.6\leq \nu \leq 1.2$) and interlayer separations ($0\leq d/\ell
\lesssim 1.0$). The phase diagram depends sensitively on the interplay of
the different symmetry breaking fields as well as filling factor and
interlayer separation. In some regions of the phase diagram, the 2DEG is
pseudospin polarized and the charged excitations are Skyrmions with real
spin texture (i.e., spin-Skyrmions) while in regions where the Zeeman
coupling dominates, the 2DEG is spin-polarized and the charged excitations
are Skyrmions with pseudospin texture (i.e., pseudospin Skyrmions or
bimerons.) The phase diagram contains also some regions where the charged
excitations involve intertwined spin and pseudospin textures. These complex
objects are called $CP^{3}$ Skyrmions\cite{rajaramancp3}. Spin and
pseudospin-Skyrmions can be viewed as limiting cases of the more general $%
CP^{3}$ Skyrmion excitation.

In the present paper, our main goal is to understand the collective mode
spectrum of a crystal of $CP^{3}$ Skyrmions. To understand the intricate
spectrum of excitations in such a crystal, we find it helpful to begin our
study by calculations of the collective mode spectra of pure spin or
pseudospin Skyrmion crystals as well as that of the homogeneous (or liquid)
state at exactly $\nu =1$.

For the liquid state, an analysis of the collective excitations was
presented in Ref. \onlinecite{hasebe}, but the analysis was limited to the
behavior at long wavelengths. We provide here the full dispersion relations,
calculated in the Generalized Random-Phase Approximation (GRPA), of the
three Goldstone modes identified in Ref. \onlinecite{hasebe}. In addition to
pseudospin-wave and spin-wave modes, the $CP^{3}$ liquid state also supports
a gapless mode involving a spin \textit{and} a pseudospin flip.

The collective mode spectrum of a pure spin-Skyrmion crystal was first
calculated in Ref.\onlinecite{cotegirvinprl}. In this work we present and
discuss both two-component cases:\ the spin and pseudospin crystals, and
discuss the similarities and differences. The spin-Skyrmion has two
Goldstone modes at finite Zeeman coupling. One is the phonon mode which is
due to the broken translational symmetry, while the second mode is a $%
U\left( 1\right) $ phase mode arising from a broken orientational symmetry
in the spin texture. Indeed, the energy of an isolated spin-Skyrmion is
invariant with respect to a global change in the azimuthal angle of the
spins. In a crystal, this angle is fixed and this symmetry is broken. It is
believed\cite{cotegirvinprl} that this mode is responsible for the small NMR
relaxation time $T_{1}$ observed experimentally\cite{barret} around $\nu =1$
at low temperature. A particularly interesting situation arises if one
considers the limit of vanishing Zeeman coupling $\Delta _{Z}$: each
Skyrmion breaks up into two merons (a meron is a vortex-like configuration
which can be understood as half a Skyrmion), and the resulting meron crystal
has $three$ Goldstone modes in addition to the phonon mode which are similar
in nature to the $U\left( 1\right) $ phase mode of the spin-Skyrmion crystal.

For the analogous bilayer situation, where the two layers in which the
electrons may reside play the role of the two allowed spin states, we find
that only the phonon mode remains gapless in the pseudospin-Skyrmion crystal
when the tunneling and layer separation are both non-vanishing. The absence
of a gapless mode analogous to the $U(1)$ mode of the spin-Skyrmion crystal
occurs because the inter-layer interaction is different than the intra-layer
interaction, breaking the $U(1)$ symmetry. For $\Delta _{SAS}\rightarrow 0$,
the state deforms into a pseudospin meron crystal, which has two (rather
than four) gapless modes because of the explicit symmetry-breaking in the
interaction. These modes may be identified as a phonon, and a Goldstone mode
due to the spontaneous interlayer coherence in the limit of zero tunneling
and the broken orientational symmetry in the in-plane pseudospin texture.

For the $CP^{3}$ Skyrmion crystal, we show that, when all symmetry breaking
fields $\Delta _{SAS}$, $\Delta _{Z},$ and $\Delta _{b}$ are finite, the
crystal has, in addition to the phonon mode, one gapless phase $U\left(
1\right) $ mode arising from the invariance of the Hamiltonian with respect
to a rotation of \textit{all} the spins in the two layers by the same angle.
We find also that since the $CP^{3}$ Skyrmion crystal occurs only at very
small value of the tunneling parameter, a second phase mode with a very
small gap is present in the spectrum. Our $CP^{3}$ Skyrmion crystal is
characterized by a set of eight low-energy modes that offer a distinct
signature of this entangled spin and pseudospin state, much different from
that of a pure spin or pseudospin Skyrmion crystal.

Our paper is organized in the following way. We introduce our description of
the crystal phases in the bilayer quantum Hall system in Sec. II and the
formalism we use to compute the dispersion relations of the collective modes
in the GRPA in Sec. III. We study the collective mode spectrum of the liquid
state for $\nu =1$ in Sec. IV, that of the pure spin-Skyrmion crystal in
Sec. V, and that of the pure pseudospin crystal in Sec. VI. We are then
ready in Sec. VII to consider the more general case of a $CP^{3}$ Skyrmion
crystal. We study the excitation spectrum of this crystal as a function of
interlayer separation, tunnel coupling and electrical bias. Finally, we
conclude in Sec. VIII.

\section{Description of the crystal states}

We consider a 2DEG\ in a double-quantum-well system (DQWS) in a quantizing
magnetic field $\mathbf{B}=B\widehat{\mathbf{z}}$ taking into account the
possibility that an electrical bias is applied to the system. We define this
bias as $\Delta _{b}=E_{R}-E_{L}$ where $E_{R\left( L\right) }$ are the
electric subband energies in each well (right and left) in the absence of
magnetic field and tunneling. For simplicity, we make a narrow well
approximation, i.e. we assume that the width $b$ of the wells is small ($%
b<<\ell $ ) and treat interlayer hopping in a tight-binding approximation.
The single-particle problem is then characterized by the total filling
factor $\nu $, the separation $d$ (from center to center) between the wells,
the tunneling gap $\Delta _{SAS}$, the electrical bias $\Delta _{b}$ and the
Zeeman energy $\Delta _{Z}=g^{\ast }\mu _{B}B$ where $g^{\ast }$ is the
effective gyromagnetic factor and $\mu _{B}$ is the Bohr magneton.

We describe the phases of the electrons in the DQWS by the set of average
fields $\left\{ \left\langle \rho _{i,j}^{\alpha ,\beta }(\mathbf{q}%
)\right\rangle \right\} $ where $\rho _{i,j}^{\alpha ,\beta }(\mathbf{q})$
is an operator that we define\cite{cotemethode} as 
\begin{equation}
\rho _{i,j}^{\alpha ,\beta }(\mathbf{q})=\frac{1}{N_{\varphi }}%
\sum_{X}e^{-iq_{x}X+iq_{x}q_{y}\ell ^{2}/2}\ c_{i,\alpha ,X}^{\dagger
}c_{j,\beta ,X-q_{y}\ell ^{2}}.
\end{equation}%
In this equation, $N_{\varphi }=S/2\pi \ell ^{2}$ is the Landau level
degeneracy with $S$ the area of the 2DEG. The indices $i,j=R,L$ and $\alpha
,\beta =\pm 1$ are respectively well and spin indices. We use the Landau
gauge with vector potential $\mathbf{A}=\left( 0,Bx,0\right) $. The operator 
$c_{i,\alpha ,X}^{\dagger }$ creates an electron in state $\left( i,\alpha
,X\right) $ in the lowest Landau level. We work in the strong magnetic field
limit where the Hilbert space is restricted to the lowest Landau level only
and make the usual approximation of neglecting Landau level mixing. For a
crystal state, only the fields $\left\langle \rho _{i,j}^{\alpha ,\beta }(%
\mathbf{q=G})\right\rangle \mathbf{\neq 0,}$ where $\mathbf{G}$ is a
reciprocal lattice vector of the crystal structure. The calculation of the
order parameters $\left\langle \rho _{i,j}^{\alpha ,\beta }(\mathbf{q=G}%
)\right\rangle $ in the Hartree-Fock approximation for a crystal of $CP^{3}$
Skyrmions was explained in Paper I\cite{cotecp3}.

The Hartree-Fock Hamiltonian of a general crystal state is given by

\begin{eqnarray}
H_{HF} &=&N_{\phi }\sum_{i,\alpha }\widetilde{E}_{\alpha ,i}\rho
_{i,i}^{\alpha ,\alpha }\left( 0\right)  \label{gsenergy} \\
&&-N_{\phi }\frac{\Delta _{SAS}}{2}\sum_{\alpha }\left( \rho _{R,L}^{\alpha
,\alpha }\left( 0\right) +\rho _{L,R}^{\alpha ,\alpha }\left( 0\right)
\right)  \notag \\
&&+N_{\phi }\sum_{\alpha ,\beta }\sum_{i,j}\sum_{\mathbf{G}}H_{i,j}\left(
G\right) \left\langle \rho _{i,i}^{\alpha ,\alpha }(-\mathbf{G}%
)\right\rangle \rho _{j,j}^{\beta ,\beta }(\mathbf{G})  \notag \\
&&-N_{\phi }\sum_{\alpha ,\beta }\sum_{i,j}\sum_{\mathbf{G}}X_{i,j}\left(
G\right) \left\langle \rho _{i,j}^{\alpha ,\beta }(-\mathbf{G})\right\rangle
\rho _{j,i}^{\beta ,\alpha }(\mathbf{G}),  \notag
\end{eqnarray}%
with the renormalized single-particle energies given by%
\begin{eqnarray}
\widetilde{E}_{R,\alpha } &=&\frac{\Delta _{b}}{2}-\alpha \frac{\Delta _{Z}}{%
2}+\left( \frac{e^{2}}{\kappa \ell }\right) \left( \frac{\nu }{2}-\nu
_{L}\right) \left( \frac{d}{\ell }\right) , \\
\widetilde{E}_{L,\alpha } &=&-\frac{\Delta _{b}}{2}-\alpha \frac{\Delta _{Z}%
}{2}+\left( \frac{e^{2}}{\kappa \ell }\right) \left( \frac{\nu }{2}-\nu
_{R}\right) \left( \frac{d}{\ell }\right) ,
\end{eqnarray}%
and the Hartree and Fock intrawell and interwell interactions defined by%
\begin{eqnarray}
H_{i,i}\left( q\right) &=&H\left( q\right) =\left( \frac{e^{2}}{\kappa \ell }%
\right) \frac{1}{q\ell }e^{-q^{2}\ell ^{2}}[1-\delta _{q,0}],  \label{h1} \\
H_{i\neq j}\left( q\right) &=&\widetilde{H}\left( q\right) =\left( \frac{%
e^{2}}{\kappa \ell }\right) \frac{1}{q\ell }e^{-q^{2}\ell
^{2}}[e^{-qd}-\delta _{q,0}],  \label{h2} \\
X_{i,i}\left( q\right) &=&X\left( q\right) =\left( \frac{e^{2}}{\kappa \ell }%
\right) \int_{0}^{+\infty }dye^{-y^{2}/2}J_{0}\left( q\ell y\right) ,
\label{f1} \\
X_{i\neq j}\left( q\right) &=&\widetilde{X}\left( q\right)  \label{f2} \\
&=&\left( \frac{e^{2}}{\kappa \ell }\right) \int_{0}^{+\infty
}dye^{-y^{2}/2}e^{-dy/\ell }J_{0}\left( q\ell y\right) .  \notag
\end{eqnarray}%
Note that to take into account a uniform neutralizing positive background,
one sets $H\left( 0\right) =0$ and $\widetilde{H}\left( 0\right) =-\left( 
\frac{e^{2}}{\kappa \ell }\right) \frac{d}{\ell }.$

For the full four-level system, an electronic state may be specified by the
four-component spinor%
\begin{equation}
c_{X}=\left( 
\begin{array}{c}
c_{R,+,X} \\ 
c_{R,-,X} \\ 
c_{L,+,X} \\ 
c_{L,-,X}%
\end{array}%
\right) .  \label{spinor}
\end{equation}%
At this point, it is convenient to redefine the operators $\rho
_{i,j}^{\alpha ,\beta }(\mathbf{q})$ introduced above as $\rho _{a,b}(%
\mathbf{q})$ with the indices $a,b$ taking the four values $a,b=1,2,3,4$ and
referring to the states $\left( R,+\right) ,(R,-),(L,+),(L,-)$ in this
order. The layer index can be mapped into a pseudospin index $\pm _{p}$with
the correspondance $R\rightarrow +_{p}$ and $L\rightarrow -_{p}$. The
density, spin density, and pseudospin density operators $\rho \left( \mathbf{%
q}\right) ,\mathbf{S}\left( \mathbf{q}\right) $ and $\mathbf{P}\left( 
\mathbf{q}\right) $ and the nine operators $R_{u,v}\left( \mathbf{q}\right) $
(with $u,v=x,y,z$) are related\cite{ezawasu4} to the four-component spinor
of Eq. (\ref{spinor}) by the equations\cite{correction} 
\begin{equation}
\rho (\mathbf{q})=\frac{1}{N_{\varphi }}\sum_{X}e^{-iq_{x}X+iq_{x}q_{y}\ell
^{2}/2}\ c_{X}^{\dagger }c_{X-q_{y}\ell ^{2}},  \label{a1}
\end{equation}%
\begin{equation}
S_{u}(\mathbf{q})=\frac{1}{2N_{\varphi }}\sum_{X}e^{-iq_{x}X+iq_{x}q_{y}\ell
^{2}/2}\ c_{X}^{\dagger }\tau _{u}^{spin}c_{X-q_{y}\ell ^{2}},  \label{a2}
\end{equation}%
\begin{equation}
P_{u}(\mathbf{q})=\frac{1}{2N_{\varphi }}\sum_{X}e^{-iq_{x}X+iq_{x}q_{y}\ell
^{2}/2}\ c_{X}^{\dagger }\tau _{u}^{ppin}c_{X-q_{y}\ell ^{2}},  \label{a3}
\end{equation}%
\begin{equation}
R_{u,v}(\mathbf{q})=\frac{1}{2N_{\varphi }}\sum_{X}e^{-iq_{x}X+iq_{x}q_{y}%
\ell ^{2}/2}\ c_{X}^{\dagger }\tau _{u}^{spin}\tau _{v}^{ppin}c_{X-q_{y}\ell
^{2}},  \label{a4}
\end{equation}%
where the $4\times 4$ matrices $\tau _{u}^{spin}$ and $\tau _{v}^{ppin}$ are
defined by

\begin{equation}
\tau _{u}^{spin}=\left( 
\begin{array}{cc}
\sigma _{u} & 0 \\ 
0 & \sigma _{u}%
\end{array}%
\right) ,
\end{equation}%
(with $\sigma _{u}$ a Pauli matrix) and by%
\begin{equation}
\tau _{x}^{ppin}=\left( 
\begin{array}{cc}
0 & I \\ 
I & 0%
\end{array}%
\right) ,\tau _{y}^{ppin}=\left( 
\begin{array}{cc}
0 & -iI \\ 
iI & 0%
\end{array}%
\right) ,
\end{equation}%
and%
\begin{equation}
\tau _{z}^{ppin}=\left( 
\begin{array}{cc}
I & 0 \\ 
0 & -I%
\end{array}%
\right) ,
\end{equation}%
where $I$ is the $2\times 2$ unit matrix.

One may show\cite{cotecp3} that the order parameters obey the four sum rules%
\begin{equation}
\sum_{\mathbf{G}}\sum_{b}\left\vert \left\langle \rho _{a,b}(\mathbf{G}%
)\right\rangle \right\vert ^{2}=\nu _{a}.  \label{sumrule}
\end{equation}

\section{Calculation of the response functions}

We define the two-particle Matsubara Green's functions as%
\begin{equation}
\chi _{a,b,c,d}\left( \mathbf{q},\mathbf{q}^{\prime };\tau \right) =-N_{\phi
}\left\langle T_{\tau }\delta \rho _{a,b}\left( \mathbf{q,}\tau \right)
\delta \rho _{c,d}\left( \mathbf{-q}^{\prime },0\right) \right\rangle ,
\label{response}
\end{equation}%
where $T_{\tau }$ is the imaginary time ordering operator. In a crystal,
these functions are non zero only for wavectors $\mathbf{q=k+G}$ and $%
\mathbf{q}^{\prime }\mathbf{=k+G}^{\prime }$ where $\mathbf{k}$ is a vector
in the first Brillouin zone of the lattice. The collective mode spectrum is
found by tracking the poles of the response functions $\chi _{a,b,c,d}\left( 
\mathbf{k+G},\mathbf{k+G};i\Omega _{n}\rightarrow \omega +i\delta \right) $
when $\mathbf{k}$ varies in the Brillouin zone.

Let us denote by $a_{l},a_{s}$ the layer and spin indices in $a$ and
similarly for the other indices $b,c,d$. The equations of motion for the
two-particle Matsubara Green's functions in the Hartree-Fock approximation
(i.e. $\chi _{a,b,c,d}^{0}$) are given by 
\begin{eqnarray}
&&\hslash \frac{\partial }{\partial \tau }\chi _{a,b,c,d}^{0}\left( \mathbf{q%
},\mathbf{q}^{\prime };\tau \right)  \label{hfequation} \\
&=&-N_{\varphi }\hslash \left\langle \left[ \rho _{a,b}\left( \mathbf{q,}%
\tau \right) ,\rho _{c,d}\left( \mathbf{-q}^{\prime },0\right) \right]
\right\rangle \delta \left( \tau \right)  \notag \\
&&-N_{\varphi }\left\langle T_{\tau }\left[ H_{HF}-\mu N,\rho _{a,b}\left( 
\mathbf{q,}\tau \right) \right] ,\delta \rho _{c,d}\left( \mathbf{-q}%
^{\prime },0\right) \right\rangle ,  \notag
\end{eqnarray}%
with $H_{HF}$ given in Eq. (\ref{gsenergy}). Evaluating the commutators and
Fourier transforming with respect to the imaginary time $\tau $, we get the
lengthy equation%
\begin{eqnarray}
&&\left[ i\hslash \Omega _{n}+\left( \widetilde{E}_{a}-\widetilde{E}%
_{b}\right) \right] \chi _{a,b,c,d}^{0}\left( \mathbf{q},\mathbf{q}^{\prime
};\Omega _{n}\right) \\
&=&\hslash \left\langle \rho _{a,d}\left( \mathbf{q-q}^{\prime }\right)
\right\rangle \delta _{b,c}\gamma _{\mathbf{q},\mathbf{q}^{\prime }}  \notag
\\
&&-\hslash \left\langle \rho _{c,b}\left( \mathbf{q-q}^{\prime }\right)
\right\rangle \delta _{a,d}\gamma _{\mathbf{q},\mathbf{q}^{\prime }}^{\ast }
\notag \\
&&+\frac{1}{2}\Delta _{SAS}\delta _{a_{l},L}\chi _{\left( a_{s},R\right)
,b,c,d}^{0}\left( \mathbf{q},\mathbf{q}^{\prime };\Omega _{n}\right)  \notag
\\
&&-\frac{1}{2}\Delta _{SAS}\delta _{b_{l},R}\chi _{a,\left( b_{s},L\right)
,c,d}^{0}\left( \mathbf{q},\mathbf{q}^{\prime };\Omega _{n}\right)  \notag \\
&&+\frac{1}{2}\Delta _{SAS}\delta _{a_{l},R}\chi _{\left( a_{s},L\right)
,b,c,d}^{0}\left( \mathbf{q},\mathbf{q}^{\prime };\Omega _{n}\right)  \notag
\\
&&-\frac{1}{2}\Delta _{SAS}\delta _{b_{l},L}\chi _{a,\left( b_{s},R\right)
,c,d}^{0}\left( \mathbf{q},\mathbf{q}^{\prime };\Omega _{n}\right)  \notag \\
&&-\sum_{a^{\prime }}\sum_{\mathbf{q}^{\prime \prime }\neq \mathbf{q}%
}U_{a^{\prime },a}^{H}\left( \mathbf{q-q}^{\prime \prime }\right) \gamma _{%
\mathbf{q},\mathbf{q}^{\prime \prime }}\chi _{a,b,c,d}^{0}\left( \mathbf{q}%
^{\prime \prime },\mathbf{q}^{\prime };\Omega _{n}\right)  \notag \\
&&+\sum_{a^{\prime }}\sum_{\mathbf{q}^{\prime \prime }\neq \mathbf{q}%
}U_{a^{\prime },b}^{H}\left( \mathbf{q-q}^{\prime \prime }\right) \gamma _{%
\mathbf{q},\mathbf{q}^{\prime \prime }}^{\ast }\chi _{a,b,c,d}^{0}\left( 
\mathbf{q}^{\prime \prime },\mathbf{q}^{\prime };\Omega _{n}\right)  \notag
\\
&&+\sum_{b^{\prime }}\sum_{\mathbf{q}^{\prime \prime }}U_{a,b^{\prime
}}^{F}\left( \mathbf{q-q}^{\prime \prime }\right) \gamma _{\mathbf{q},%
\mathbf{q}^{\prime \prime }}\chi _{b^{\prime },b,c,d}^{0}\left( \mathbf{q}%
^{\prime \prime },\mathbf{q}^{\prime };\Omega _{n}\right)  \notag \\
&&-\sum_{a^{\prime }}\sum_{\mathbf{q}^{\prime \prime }}U_{a^{\prime
},b}^{F}\left( \mathbf{q-q}^{\prime \prime }\right) \gamma _{\mathbf{q},%
\mathbf{q}^{\prime \prime }}^{\ast }\chi _{a,a^{\prime }c,d}^{0}\left( 
\mathbf{q}^{\prime \prime },\mathbf{q}^{\prime };\Omega _{n}\right) ,  \notag
\end{eqnarray}%
where $\gamma _{\mathbf{q},\mathbf{q}^{\prime }}=e^{i\left( \mathbf{q}\times 
\mathbf{q}^{\prime }\right) \cdot \widehat{\mathbf{z}}\ell ^{2}/2}$ and we
have defined the mean-field Hartree and Fock potentials%
\begin{eqnarray}
U_{a,d}^{H}\left( \mathbf{q}\right) &=&H_{a_{l},d_{l}}\left( \mathbf{q}%
\right) \left\langle \rho ^{a,a}\left( \mathbf{q}\right) \right\rangle , \\
U_{a,d}^{F}\left( \mathbf{q}\right) &=&X_{a_{l},d_{l}}\left( \mathbf{q}%
\right) \left\langle \rho ^{a,d}\left( \mathbf{q}\right) \right\rangle ,
\end{eqnarray}%
with the Hartree and Fock interactions defined in Eqs. (\ref{h1}-\ref{f2}).

We get the response functions in the GRPA\ by summing the ladder and bubble
diagrams\cite{cotemethode}. The result is the integral equation 
\begin{align}
& \chi _{a,b,c,d}\left( \mathbf{q},\mathbf{q}^{\prime };\Omega _{n}\right)
\label{grpa_1} \\
& =\chi _{a,b,c,d}^{0}\left( \mathbf{q},\mathbf{q}^{\prime };\Omega
_{n}\right)  \notag \\
& +\frac{1}{\hslash }\sum_{\mathbf{q}^{\prime \prime }}\chi
_{a,b,e,e}^{0}\left( \mathbf{q},\mathbf{q}^{\prime \prime };\Omega
_{n}\right) H_{e,g}\left( \mathbf{q}^{\prime \prime }\right) \chi
_{g,g,c,d}\left( \mathbf{q}^{\prime \prime },\mathbf{q}^{\prime };\Omega
_{n}\right)  \notag \\
& -\frac{1}{\hslash }\sum_{\mathbf{q}^{\prime \prime }}\chi
_{a,b,e,f}^{0}\left( \mathbf{q},\mathbf{q}^{\prime \prime };\Omega
_{n}\right) X_{e,f}\left( \mathbf{q}^{\prime \prime }\right) \chi
_{f,e,c,d}\left( \mathbf{q}^{\prime \prime },\mathbf{q}^{\prime };\Omega
_{n}\right) .  \notag
\end{align}

We now write Eq. (\ref{grpa_1}) in a matrix form more appropriate for
numerical analysis. We first redefine $\chi _{a,b,c,d}$ as the matrix $\chi
_{i,j}$ with the indices $i,j$ taking the values from $1$ to $16$ with the
order: $(1,1),(1,2),(1,3),(1,4),(2,1),$\textit{etc}. for $i,$ the row index,
and the order: $(1,1),(2,1),(3,1),(4,1),(1,2),$ \textit{etc}. for $j$, the
column index. It is to be noted that all elements $\chi _{i,j}$ of the
matrix $\chi $ as well as of the other matrices $I,A,B,H,X$ that we define
below are themselves matrices in the reciprocal lattice vectors $\mathbf{G},%
\mathbf{G}^{\prime }.$ In the end, we have to diagonalize a matrix with
dimensions $16n_{RLV}\times 16n_{RLV}$ where $n_{RLV}$ is the number of
reciprocal lattice vectors $\mathbf{G}$ that we keep in the calculation. We
typically work with $n_{RLV}\approx 317$ (in the calculation of the order
parameters, however, we keep more than $600$ reciprocal lattice vectors). We
then write the index $i$ as $i=4\left( \overline{i}-1\right) +\widetilde{i}$
where $\overline{i},\widetilde{i}=1,2,3,4$ and similarly for $j$. The unit
matrix is written as%
\begin{equation}
I=\delta _{\overline{i},\overline{j}}\delta _{\widetilde{i},\widetilde{j}%
}\delta _{\mathbf{q,q}^{\prime }}.
\end{equation}%
We define the following matrices:%
\begin{eqnarray}
B_{i,j}\left( \mathbf{q},\mathbf{q}^{\prime }\right) &=&\left\langle \rho _{%
\overline{i},\overline{j}}\left( \mathbf{q-q}^{\prime }\right) \right\rangle
\gamma _{\mathbf{q},\mathbf{q}^{\prime }}\delta _{\widetilde{i},\widetilde{j}%
} \\
&&-\left\langle \rho _{\widetilde{j},\widetilde{i}}\left( \mathbf{q-q}%
^{\prime }\right) \right\rangle \gamma _{\mathbf{q},\mathbf{q}^{\prime
}}^{\ast }\delta _{\overline{i},\overline{j}},  \notag
\end{eqnarray}%
\begin{equation}
E_{i,j}\left( \mathbf{q},\mathbf{q}^{\prime }\right) =-\left( \widetilde{E}_{%
\overline{i}}-\widetilde{E}_{\widetilde{i}}\right) \delta _{i,j}\delta _{%
\mathbf{q,q}^{\prime }},
\end{equation}%
\begin{eqnarray}
F_{i,j}\left( \mathbf{q},\mathbf{q}^{\prime }\right) &=&-\left( H_{\overline{%
i}}^{+}\left( \mathbf{q},\mathbf{q}^{\prime }\right) -H_{\widetilde{i}%
}^{-}\left( \mathbf{q},\mathbf{q}^{\prime }\right) \right) \delta _{i,j} \\
&&+F_{\overline{i},\overline{j}}^{+}\left( \mathbf{q},\mathbf{q}^{\prime
}\right) \delta _{\widetilde{i},\widetilde{j}}-F_{\widetilde{j},\widetilde{i}%
}^{-}\left( \mathbf{q},\mathbf{q}^{\prime }\right) \delta _{\overline{i},%
\overline{j}},  \notag
\end{eqnarray}%
\begin{eqnarray}
T_{i,j}\left( \mathbf{q},\mathbf{q}^{\prime }\right) &=&\frac{\Delta _{SAS}}{%
2}\delta _{\overline{i},\overline{j}}\delta _{\left\vert \widetilde{i}-%
\widetilde{j}\right\vert ,2}\delta _{\mathbf{q,q}^{\prime }}, \\
&&-\frac{\Delta _{SAS}}{2}\delta _{\widetilde{i},\widetilde{j}}\delta
_{\left\vert \overline{i}-\overline{j}\right\vert ,2}\delta _{\mathbf{q,q}%
^{\prime }}  \notag
\end{eqnarray}%
with the definitions (with $k,l=1,2,3,4$)%
\begin{equation}
H_{k}^{\pm }\left( \mathbf{q},\mathbf{q}^{\prime }\right) =\sum_{a^{\prime
}}U_{a^{\prime },k}^{H}\left( \mathbf{q-q}^{\prime }\right) e^{\pm i\left( 
\mathbf{q}\times \mathbf{q}^{\prime \prime }\right) \cdot \widehat{\mathbf{z}%
}\ell ^{2}/2},
\end{equation}%
and%
\begin{equation*}
F_{k,l}^{\pm }\left( \mathbf{q-q}^{\prime \prime }\right) =U_{k,l}^{F}\left( 
\mathbf{q-q}^{\prime }\right) e^{\pm i\left( \mathbf{q}\times \mathbf{q}%
^{\prime \prime }\right) \cdot \widehat{\mathbf{z}}\ell ^{2}/2}.
\end{equation*}%
We also define the interactions%
\begin{equation}
V_{i,j}\left( \mathbf{q},\mathbf{q}^{\prime }\right) =\left[ H_{\overline{i},%
\overline{j}}\left( \mathbf{q}\right) \delta _{\widetilde{i},\overline{i}%
}\delta _{\widetilde{j},\overline{j}}-X_{\widetilde{j},\overline{j}}\left( 
\mathbf{q}\right) \delta _{i,j}\right] \delta _{\mathbf{q,q}^{\prime }}.
\end{equation}%
With all these definitions, we can write the equation of motion for the
response functions in the matrix form:

\begin{equation}
\left[ i\hslash \Omega _{n}I-E-T-F-BV\right] \chi =\hslash B.  \label{grpa}
\end{equation}%
To get the weight of each pole, we diagonalize the matrix $M=E+T+F+BV$ by
writing $MU=UD$ where $U$ is the matrix containing the eigenvectors and $D$
is the diagonal matrix of the eigenvalues $\hslash \omega _{j}\left( \mathbf{%
k}\right) $. The response function $\chi $ can be written as%
\begin{equation}
\chi _{n,m}\left( \mathbf{k},\omega \right) =\sum_{j,k}\frac{U_{n,j}\left( 
\mathbf{k}\right) \left[ U\left( \mathbf{k}\right) \right]
_{j,k}^{-1}B_{k,m}\left( \mathbf{k}\right) }{\omega +i\delta -\omega
_{j}\left( \mathbf{k}\right) }
\end{equation}%
where the indices $n,m,j,k=1,2,...,16n_{RLV}.$ Using Eqs. (\ref{a1}-\ref{a4}%
), it is easy to get from $\chi $ the response functions for the density,
spin, pseudospin and the $R$ fields. Because we have access to the matrix of
eigenvectors $U$, we can also animate the motion of the density, spin,
pseudospin, etc. in a given mode. This will be very helpful for the
interpretation of the different modes.

\section{Uniform coherent state at $\protect\nu =1$}

In this section, we apply our formalism to the study of the spin-polarized
uniform coherent state (UCS) at $\nu =1$. For effectively spinless
electrons, the behavior of the collective mode (i.e. pseudospin wave) with
bias and interlayer separation was studied before in Ref. %
\onlinecite{joglekar}. One may also consider\cite{hasebe} the limit of an $%
SU(4)\ $ symmetric Hamiltonian, in which the interlayer separation $d$, $%
\Delta _{Z}$, and $\Delta _{SAS}$ are all set to zero. In this case three
Goldstone modes were predicted in the absence of symmetry-breaking fields%
\cite{hasebe} using a gradient expansion, but quantitative dispersion
relations for these modes that takes into account the spin, pseudospin and
density fields are difficult to obtain in this approach. Our GRPA\ formalism
produces these three modes, provides analytical results for the dispersions
in some limiting cases, and allows numerical results to be obtained in the
general case from the microscopic Hamiltonian.

In UCS, the only non-zero order parameters are 
\begin{eqnarray}
\left\langle \rho _{1,1}\left( 0\right) \right\rangle &=&\nu _{1},  \notag \\
\left\langle \rho _{3,3}\left( 0\right) \right\rangle &=&\nu _{3},
\label{orderp} \\
\left\langle \rho _{1,3}\left( 0\right) \right\rangle &=&\left\vert
\left\langle \rho _{3,1}\left( 0\right) \right\rangle \right\vert ^{\ast
}=\alpha ,  \notag
\end{eqnarray}%
(we can choose $\alpha $ real without lost of generality). For a balanced
DQWS, $\nu _{1}=\nu _{3}$ while in the presence of a bias, $\nu _{1},\nu
_{3} $ and $\alpha $ must be found by minimizing the Hartree-Fock energy. We
have discussed the effect of a bias on the UCS in Paper I\cite{cotecp3}. In
the absence of tunneling, we obtain a simple relation between charge
imbalance and bias, 
\begin{equation}
\sigma =\nu _{3}-\nu _{1}=\frac{-\Delta _{b}}{X\left( 0\right) -\widetilde{X}%
\left( 0\right) -\frac{d}{\ell }\left( \frac{e^{2}}{\kappa \ell }\right) }.
\label{bias}
\end{equation}

In the UCS, the matrices in Eq. (\ref{grpa}) have dimensions $16\times 16$
and the system of equations split into three uncoupled systems of equations
corresponding to the $16$ excitations represented schematically in Fig. 1
(transitions 1 to 4 are intra-level excitations with zero energy). The
excitations represented with the dashed lines in this figure are just the
reverse of those represented with the full lines. In this figure, we have
illustrated the situation where $\Delta _{b}=0$ and $\Delta _{SAS}<\Delta
_{Z}$ so that the first two levels are the symmetric ($S$) and antisymmetric
($AS$) states of the DQWS with real spin up. Because only the $S,+$ level is
filled in the ground state, our calculation gives three dispersive modes
corresponding to the transitions from the filled level $S,+$ numbered $5,11$
and $13$ in Fig. 1. We will not discuss the other non dispersive modes as
they have zero weight in the response functions at zero temperature. Note
that, at finite bias, the bonding and antibonding states (with
single-particle energies $\pm \sqrt{\Delta _{SAS}^{2}+\Delta _{b}^{2}}$)
replace the $S$ and $AS$ states in the diagrams of Fig. 1.

\begin{figure}[tbph]
\includegraphics[scale=.45]{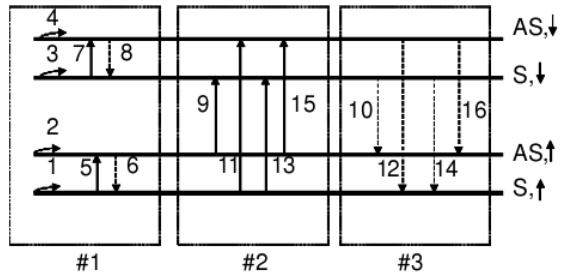}
\caption{The 16 possible transitions between the single-particle energy
levels in the UCS is divided in the GRPA\ into 3 uncoupled subsystems
involving 8,4 and 4 transitions. }
\end{figure}

Transition $5$ corresponds to a pseudospin-wave mode first found in Ref. %
\onlinecite{fertigdispersion} in which the pseudospins precess around their
groundstate orientation. Solving Eq. (\ref{grpa}), we find the general
dispersion relation 
\begin{eqnarray}
\hslash ^{2}\omega _{5}^{2}\left( q\right) &=&4a\left( q\right) b\left(
q\right)  \label{pseudowave} \\
&&+\left[ \Delta _{b}+\sigma \left( X\left( 0\right) -\widetilde{X}\left(
q\right) -\frac{d}{\ell }\left( \frac{e^{2}}{\kappa \ell }\right) \right) %
\right] ^{2},  \notag
\end{eqnarray}%
where 
\begin{eqnarray}
a\left( q\right) &=&\frac{\Delta _{SAS}}{2}+\alpha \left[ \widetilde{X}%
\left( 0\right) -\widetilde{X}\left( q\right) \right] , \\
b\left( q\right) &=&\frac{\Delta _{SAS}}{2}  \notag \\
&&+\alpha \left[ \widetilde{X}\left( 0\right) -X\left( q\right) +H\left(
q\right) -\widetilde{H}\left( q\right) \right] .
\end{eqnarray}%
Using Eq. (\ref{bias}), it is easy to show that this mode is gapless if $%
\allowbreak \allowbreak \Delta _{SAS}=0$ \textit{even with finite bias}.
This is true as long as the interlayer coherence is non zero i.e. $\alpha
\neq 0.$ At $q=0$, the pseudospins execute small oscillations in $x-y$ plane
if the bias is zero while, at finite bias and $q=0$, they execute small
oscillations in a plane where $\left\langle P_{z}\right\rangle =cst.$ In
both cases, the energy cost is zero because the Hamiltonian of Eq. (\ref%
{gsenergy}) is invariant with respect to a rotation in a plane of constant $%
\left\langle P_{z}\right\rangle $ if $\Delta _{SAS}=0.$ At finite tunneling,
the pseudospin-wave mode is gapped with $\Delta _{SAS}^{\ast }$, the
frequency at $q=0$, larger than $\Delta _{SAS}.$

The dispersion relations of the collective modes numbered $9,11,13,15$ are
found by solving $\det B=0$ with the matrix $B$ given by

\begin{widetext}%
\begin{equation}
B=\left(
\begin{array}{cccc}
c_{1}-\nu _{1}X\left( q\right)  & -\frac{\Delta _{SAS}}{2} & f-\alpha
\widetilde{X}\left( q\right)  & 0 \\
-\frac{\Delta _{SAS}}{2} & c_{1}-\nu _{1}\widetilde{X}\left( q\right) -%
\widetilde{\Delta }_{b} & 0 & f-\alpha X\left( q\right)  \\
f-\alpha X\left( q\right)  & 0 & c_{3}-\nu _{3}\widetilde{X}\left( q\right) +%
\widetilde{\Delta }_{b} & -\frac{\Delta _{SAS}}{2} \\
0 & f-\alpha \widetilde{X}\left( q\right)  & -\frac{\Delta _{SAS}}{2} &
c_{3}-\nu _{3}X\left( q\right)
\end{array}%
\right) .
\label{fprime}
\end{equation}

\end{widetext}$\allowbreak $ In this equation, we have defined the constants%
\begin{equation}
c_{1}=\Delta _{Z}+\nu _{1}X\left( 0\right) ,
\end{equation}%
\begin{equation}
c_{3}=\Delta _{Z}+\nu _{3}X\left( 0\right) ,
\end{equation}%
\begin{equation}
f=\frac{\Delta _{SAS}}{2}+\alpha \widetilde{X}\left( 0\right) ,
\end{equation}%
\begin{equation}
\widetilde{\Delta }_{b}=\Delta _{b}+\left( \nu _{3}-\nu _{1}\right) \frac{d}{%
\ell }\left( \frac{e^{2}}{\kappa \ell }\right) .
\end{equation}%
For zero bias (in which case $\alpha =1/2$), we can solve analytically to
find the two dispersive modes%
\begin{eqnarray}
\hslash \omega _{11}\left( q\right) &=&\Delta _{Z}+\Delta _{SAS} \\
&&+\frac{1}{2}\left[ X\left( 0\right) +\widetilde{X}\left( 0\right) -X\left(
q\right) -\widetilde{X}\left( q\right) \right] ,  \notag \\
\hslash \omega _{13}\left( q\right) &=&\Delta _{Z} \\
&&+\frac{1}{2}\left[ X\left( 0\right) +\widetilde{X}\left( 0\right) -X\left(
q\right) -\widetilde{X}\left( q\right) \right] .  \notag
\end{eqnarray}%
The frequency $\omega _{13}\left( q\right) $ corresponds to the spin-wave
(or Zeeman) mode in which the spins precess around their equilibrium
orientations while mode $11$ involves a spin \textit{and} pseudospin flip.
At zero bias and $\Delta _{SAS}=0$, these two modes are degenerate at $q=0$
i.e. $\hslash \omega _{11}\left( 0\right) =\hslash \omega _{13}\left(
0\right) =\Delta _{Z}.$

Fig. 2 shows the evolution of the dispersion of the three modes $5,11,13$
with bias at $d/\ell =1,\Delta _{SAS}=0$, $\Delta _{Z}/\left( e^{2}/\kappa
\ell \right) =0.2$ for (a) $\Delta _{b}/\left( e^{2}/\kappa \ell \right) =0,$
(b) $\Delta _{b}/\left( e^{2}/\kappa \ell \right) =0.1,$ and (c) $\Delta
_{b}/\left( e^{2}/\kappa \ell \right) =0.4.$ For case (d), $\Delta
_{b}/\left( e^{2}/\kappa \ell \right) =0.4$ and $\Delta _{SAS}/\left(
e^{2}/\kappa \ell \right) =0.02$. Our numerical method also produces the
frequency of the non-dispersive transitions $7,9,15$. We are not interested
in these transitions as they have zero weigth in the response functions at
zero temprature.

From Fig. 2, we see that there is one pseudospin mode and two spin-wave
modes. We find that all three dispersions are affected by the bias. Figure
2(a) shows a roton minimum $\Delta _{b}/\left( e^{2}/\kappa \ell \right)
=0.4 $ increasing with interlayer separation indicating the instability of
the UCS. Figures (b) and (c), show that an increase in the bias removes this
roton mimimum.

The value $\Delta _{b}/\left( e^{2}/\kappa \ell \right) =0.2$ in Fig. 2(b)
corresponds to the filling factors $\nu _{1}=0.38,\nu _{3}=0.62$ while the
value $\Delta _{b}/\left( e^{2}/\kappa \ell \right) =0.4023$ in Fig. 2(c)
corresponds to $\nu _{1}=0.00008,\nu _{3}=0.99992.$ Thus, the bias in Fig.
2(c) is just below the critical value above which the charge is completely
transferred to the left well. It is easy to get analytical results for the
asymptotic values ($q\rightarrow \infty $) of the modes in this case (i.e.
the transport gaps). We need only solve Eq. (\ref{fprime}) with $t=0,\nu
_{R}=0$ and take $X\left( \mathbf{q}\right) ,\widetilde{X}\left( \mathbf{q}%
\right) \rightarrow 0$. We find a spin-wave (SW) and a spin-wave with
pseudospin flid (SWPF) with dispersions $\hslash \omega _{SWPF}\left(
q\rightarrow \infty \right) =$ $\Delta _{b}+\Delta _{Z}-\frac{d}{\ell }%
(e^{2}/\kappa \ell )+X\left( 0\right) $ and $\hslash \omega _{SW}\left(
q\rightarrow \infty \right) =\Delta _{Z}+X\left( 0\right) .$ In this limit,
the $S$ and $AS$ symmetric states have evolved into the $L$ and $R$ states.
The mode SW corresponds to a spin flip transition in the same well and its
energy is $\hslash \omega _{SW}\left( q\rightarrow \infty \right) =\Delta
_{Z}+X\left( 0\right) $ since $X\left( 0\right) $ is the energy required to
remove an electron from the filled $L,\uparrow $ state. This mode is the
spin-wave mode in a single quantum well system. Its dispersion relation at
finite $q$ is given by%
\begin{equation}
\hslash \omega _{SW}\left( \mathbf{q}\right) =\Delta _{Z}+X\left( 0\right)
-X\left( \mathbf{q}\right) .
\end{equation}%
The mode SWPF involves a spin and a pseudospin flip and so a transition of
an electron from the $L$ to the $R$ well. When the charge is completely
transferred into the left well and we take into account the positive
neutralizing backgrounds, there is an electric field oriented from the $R$
well to the $L$ well. There is thus a gain in energy $\frac{d}{\ell }%
(e^{2}/\kappa \ell )$ in transferring the charge from the $L$ to the $R$
well. The total energy to remove an electron from the filled $L,\uparrow $
state and create an infinitely separated electron-hole pair with spin and
pseudospin flip is thus $\hslash \omega _{SWPF}\left( q\rightarrow \infty
\right) =$ $\Delta _{b}+\Delta _{Z}-\frac{d}{\ell }(e^{2}/\kappa \ell
)+X\left( 0\right) $. With the parameters of Fig. 2, we find $\hslash \omega
_{SW}\left( q\rightarrow \infty \right) =1.25(e^{2}/\kappa \ell )$ and $%
\hslash \omega _{SWPF}\left( q\rightarrow \infty \right) =0.85(e^{2}/\kappa
\ell ).$ The values are in good agreement with the numerical result plotted
in Fig. 2(c). In Fig. 2(d), we have added a small tunnel coupling $\Delta
_{SAS}/(e^{2}/\kappa \ell )=0.02$ that lifts the degeneracy of the two modes
gapped at the Zeeman energy. At $q=0$, the ordering of these two modes is
just the opposite than at $q\rightarrow \infty $. The Zeeman mode is gapped
at $\hslash \omega _{SW}\left( 0\right) =\Delta _{Z}$ while the
pseudospin+spin flip mode is gapped at $\hslash \omega _{SWPF}\left(
0\right) =\Delta _{Z}+\Delta _{SAS}.$

\begin{figure}[tbph]
\includegraphics[scale=1]{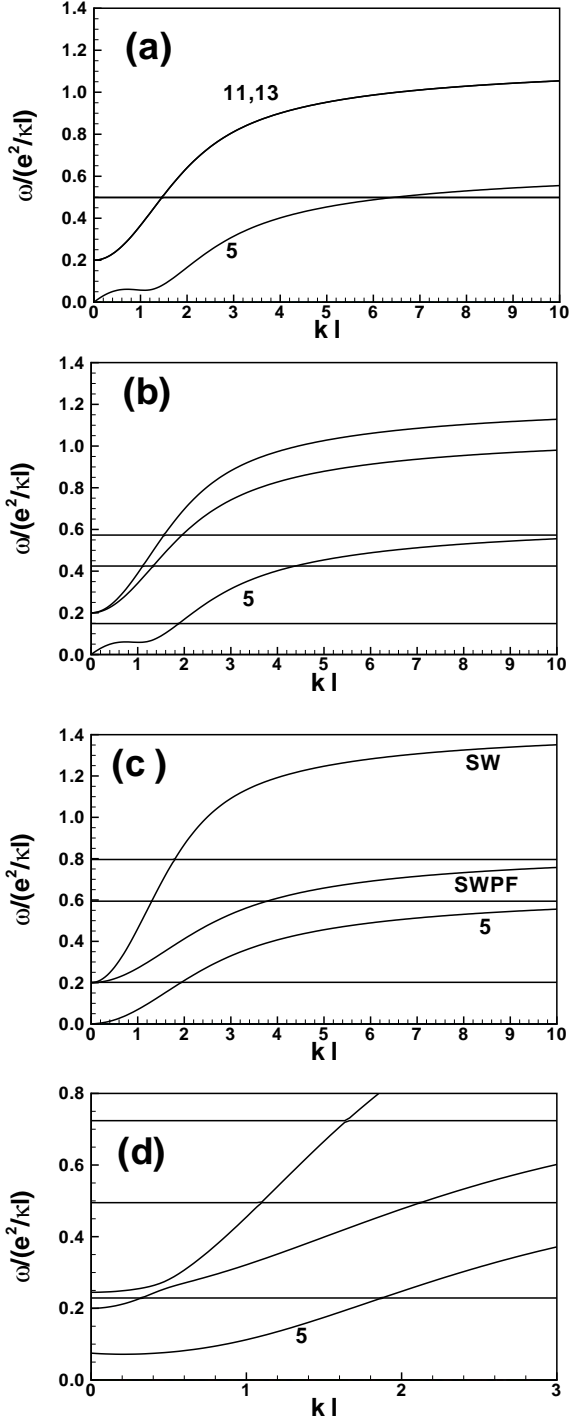}
\caption{Dispersion relations of the collective excitations in the UCS at $%
\protect\nu =1$ as computed in the GRPA. Parameters are $d/\ell =1,\Delta
_{Z}/\left( e^{2}/\protect\kappa \ell \right) =0.2$ and (a) $\Delta
_{b}/\left( e^{2}/\protect\kappa \ell \right) =0,\Delta _{SAS}=0$, (b) $%
\Delta _{b}/\left( e^{2}/\protect\kappa \ell \right) =0.1,\Delta _{SAS}=0,$
(c) $\Delta _{b}/\left( e^{2}/\protect\kappa \ell \right) =0.4023,\Delta
_{SAS}=0,$ and (d) $\Delta _{b}/\left( e^{2}/\protect\kappa \ell \right)
=0.4,$ $\Delta _{SAS}/\left( e^{2}/\protect\kappa \ell \right) =0.02.$ The
straight lines give the frequency of the non-dispersive modes. The modes $%
5,11,13$ correspond to the dispersions $\protect\omega _{5}\left( q\right) ,%
\protect\omega _{11}\left( q\right) ,\protect\omega _{13}\left( q\right) $
derived in the text. The abbreviations SW and SWPF refer to the spin-wave
and spin-wave with pseudospin-flip modes (see text).}
\end{figure}

\section{Spin-Skyrmion crystals}

If the electrical bias is sufficiently strong for all the charges to go into
the left well, the system becomes effectively a single quantum well system
(SQWS). The index $i$ then takes the values $i=3,4$ only. It is well known
that, in this case, the ground state at $\nu =1$ is a spin ferromagnet. At
small $\Delta _{Z}$, this state supports topological spin-texture
excitations known as Skyrmions\cite{sondhi,fertigskyrmion}. Away from $\nu
=1 $, a finite density $n_{s}=\left\vert \nu -1\right\vert /2\pi \ell ^{2}$
of Skyrmions is included in the the ground state and is expected to form a
Skyrme crystal\cite{breyprl}. The Hartree-Fock ground state energy of the
Skyrme crystal in a SQWS is, using Eq. (\ref{gsenergy}), 
\begin{eqnarray}
\frac{E_{HF}}{N} &=&-\frac{\Delta _{Z}}{\nu }\left\langle S_{z}\left( 
\mathbf{0}\right) \right\rangle  \label{eskyrmion} \\
&&+\frac{1}{4\nu }\sum_{\mathbf{G}}\Upsilon _{2}\left( \mathbf{G}\right)
\left\vert \left\langle \rho \left( \mathbf{G}\right) \right\rangle
\right\vert ^{2}  \notag \\
&&-\frac{1}{\nu }\sum_{\mathbf{G}}X\left( \mathbf{G}\right) \left\vert
\left\langle \mathbf{S}\left( \mathbf{G}\right) \right\rangle \right\vert
^{2},  \notag
\end{eqnarray}%
where we have defined%
\begin{equation}
\Upsilon _{2}\left( \mathbf{G}\right) =2H\left( \mathbf{G}\right) -X\left( 
\mathbf{G}\right) .
\end{equation}

A Skyrme crystal has a noncollinear magnetic order. A single Skyrmion spin
texture of such a crystal has its spins aligned with the Zeeman field at
infinity, reversed at the center of the Skyrmion, and has nonzero $XY$ spin
components at intermediate distances which have a vortex-like configuration
(see Fig. 3(a)). The classical (or quantum mean field) energy of a Skyrmion
is independent of the angle $\varphi $ which defines the global orientation
of the $XY\ $spin component. This independence gives an extra $U\left(
1\right) $ degree of freedom for a single Skyrmion. In a crystal of
Skyrmions, this $U\left( 1\right) $ symmetry is spontaneously broken and the
crystal has an extra Goldstone mode. We refer to this mode as the $XY\ $mode
because it corresponds to an oscillation of the perpendicular component of
the spins in the $x-y$ plane with $\left\langle S_{z}\right\rangle $ fixed.
From Eq. (\ref{eskyrmion}), we see that such a motion costs no energy
provided the density $\left\langle \rho \left( \mathbf{G}\right)
\right\rangle $ is not changed. That this is the case can be seen from the
fact that the relation between the topological charge and the spin texture%
\cite{leekane} is given at $\nu =1$ by (the sum over indices is implicit) 
\begin{equation}
\delta \left\langle n\left( \mathbf{r}\right) \right\rangle =-\frac{1}{8\pi }%
\varepsilon _{abc}s_{a}\left( \mathbf{r}\right) \varepsilon _{ij}\partial
_{i}s_{b}\left( \mathbf{r}\right) \partial _{j}s_{c}\left( \mathbf{r}\right)
,  \label{topolo}
\end{equation}%
where $\varepsilon _{ij}$ and $\varepsilon _{abc}$ are antisymmetric
tensors, with $i,j=x,y$ and $a,b,c=x,y,z$, and $\mathbf{s}\left( \mathbf{r}%
\right) $ is a classical field with unit modulus representing the spins. If
we write a general spin texture as

\begin{eqnarray}
s_{x}\left( \mathbf{r}\right) &=&\sin \left[ \theta \left( \mathbf{r}\right) %
\right] \cos \left[ \varphi \left( \mathbf{r}\right) \right] , \\
s_{y}\left( \mathbf{r}\right) &=&\sin \left[ \theta \left( \mathbf{r}\right) %
\right] \sin \left[ \varphi \left( \mathbf{r}\right) \right] , \\
s_{z}\left( \mathbf{r}\right) &=&\cos \left[ \theta \left( \mathbf{r}\right) %
\right] ,
\end{eqnarray}%
then the induced density takes the simple form%
\begin{equation}
\delta n\left( \mathbf{r}\right) =\frac{1}{4\pi }\sin \left[ \theta \left( 
\mathbf{r}\right) \right] \left[ \nabla \theta \left( \mathbf{r}\right)
\times \nabla \varphi \left( \mathbf{r}\right) \right] \cdot \widehat{%
\mathbf{z}}.
\end{equation}
Clearly, the induced charge density in the $XY$ mode is zero because $\nabla
\varphi \left( \mathbf{r}\right) $ is unchanged if we rotate all the spins
by the same angle (i.e. $\varphi \left( \mathbf{r}\right) \rightarrow
\varphi \left( \mathbf{r}\right) +cst$).

The phase diagram of the 2DEG around $\nu =1$ has been studied in the HFA in
Refs. \onlinecite{breyprl,cotegirvinprl,breymerons}. In a large portion of
the $\nu -\Delta _{Z}$ phase space, the Hartree-Fock ground state is a
square lattice with two Skyrmions of opposite global phase $\varphi $ per
unit cell as shown in Fig. 3(a). This configuration is called SLA (Square
Lattice Antiferromagnet) to account for the phases $\varphi =0$ and $\varphi
=\pi $ of the two Skyrmions in the unit cell. As the Zeeman energy is
increased, our Hartree-Fock approximation (HFA)\cite{cotegroup1} shows that
there is a transition to a triangular lattice of Skyrmions with a phase
difference of $\varphi =2\pi /3$ between the three Skyrmions in a unit cell
and then to a triangular Wigner crystal of spin down electrons with no spin
texture.

Note that the size of the individual Skyrmions in the Skyrme crystal is
determined in part by the Zeeman energy $\Delta _{Z}$. For large $\Delta
_{Z},$ the tipping of spins into the minority direction that occurs in the
spin texture of a skyrmion becomes energetically unfavorable, and Skyrmions
become relatively small. For decreasing $\Delta _{Z}$, the Skyrmions become
larger until their size is limited by inter-Skyrmion interactions. For small
enough Zeeman coupling the system undergoes a transition into a meron crystal%
\cite{breymerons} with four merons per unit cell as shown in Fig. 3(b). 

A meron is most easily visualized by rotating the spin axes around the $y$
axis so that the $z$ axis becomes the $x$ axis and the $x$ axis becomes the $%
-z$ axis. In this basis a meron has its spin down or up at the origin and
lies in the $XY$ plane at infinity with a vortex-like structure. The
vorticity can be positive or negative so that, combined with the up or down
spins at the origin, there are four possible types of merons. In the rotated
basis, one can recognize the Skyrmion as a bound meron-antimeron pair (i.e.,
a bimeron). Hartree-Fock calculations\cite{cotegroup1} shows that the meron
crystal is stable only at filling factors $\nu >1.38.$ For smaller filling
factors, we find, as suggested in Ref. \onlinecite{nazarov}, that Skyrmions
always appear in pairs to form a triangular Wigner lattice. There is an
attractive interaction between Skyrmions with opposite phases which prevails
over the Coulomb repulsion at large separations between the Skyrmions. Thus,
as the Zeeman energy is reduced and for $\nu _{c}<\nu <1.38$ where $\nu
_{c}\approx 1.06$, there is a transition from the Skyrme crystal to a
crystal of biskyrmions. In Ref. \onlinecite{nazarov}, the minimal energy is
found for zero separation between the pairs of Skyrmions in a biskyrmion
leading to a Skyrmion with topological charge $Q=2.$ Within the HFA, there
is a finite separation between the two Skyrmions in each pair.

\begin{figure}[tbph]
\includegraphics[scale=1]{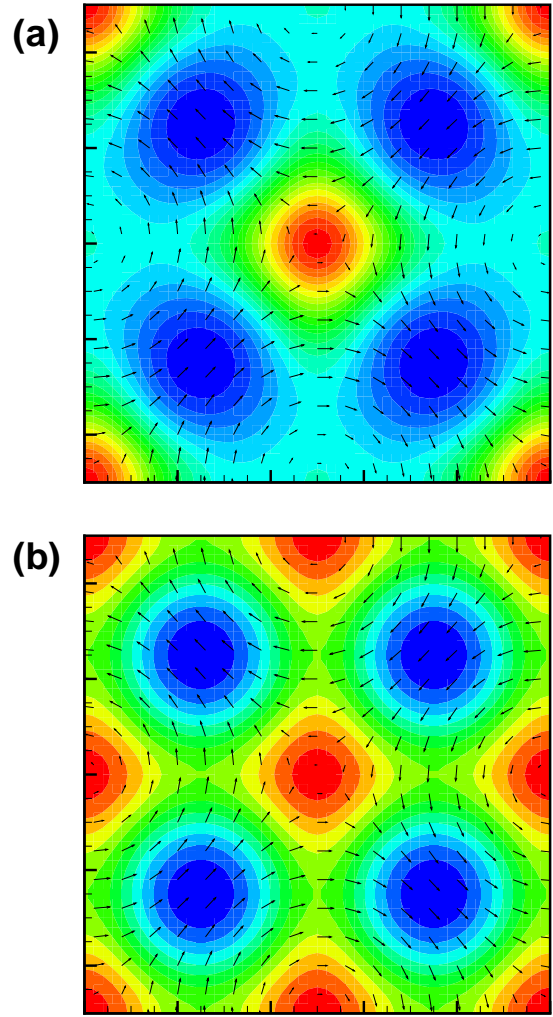}
\caption{(Color online) Spin texture in the $x-y$ plane (vectors) and
density in a unit cell of a Skyrme lattice in (a) the SLA and (b) the meron
configurations at $\protect\nu =1.30$ with $\Delta _{Z}/\left( e^{2}/\protect%
\kappa \ell \right) =0.01$ and $\Delta _{Z}/\left( e^{2}/\protect\kappa \ell
\right) =0$ respectively. In (a),\ $S_{z}=-\frac{1}{2}$ at the position of
the Skyrmions and each Skyrmion carries charge $q=-e.$ In (b), $S_{z}=-\frac{%
1}{2}$ at the positions of the four merons at the corner of the unit cell
and at the position of the meron at the center while $S_{z}=+\frac{1}{2}$ at
the positions of the other merons. Each merons carries a charge $q=-\frac{e}{%
2}.$ Contours with dark (red) regions indicate a high density.}
\end{figure}

The low-energy collective modes of the Skyrme\cite{cotegirvinprl} and meron
crystals computed in the GRPA are shown in Fig. 4. The dispersions are shown
for wavevector $\mathbf{k}$ along the path $\Gamma -M-X-\Gamma $ around the
irreducible Brillouin zone as shown in the inset of Fig. 4(b). The
low-energy part of the spectrum consists of two gapless\ (Goldstone)\ and
two gapped\ modes. One of the Goldstone mode is the phonon mode which is due
to the spontaneously broken translational symmetry in the crystal. Its
dispersion is $\omega \sim k^{3/2}$ which is similar to that of a Wigner
crystal in a magnetic field. The second Goldstone mode is the $XY$ mode
introduced above. This gapless mode is believed to be responsible\cite%
{cotegirvinprl,green} for the small spin-lattice relaxation time $T_{1}$
measured in NMR\ experiments\cite{barret} on both sides of $\nu =1$.

Of the two gapped branches, one has a gap that varies with $\Delta _{Z}$ and 
$\nu $ and the other is gapped at exactly the Zeeman energy for all $\nu $
as required by Larmor's theorem. This last mode is obviously a spin wave
mode where the spins precess around their local orientation. We refer to
this mode as the Zeeman mode. In the first mode, the motion of the two
Skyrmions with opposite global phases $\varphi $ in a unit cell (as seen in
our animations\cite{animation}) are out of phase. We identify this mode as
an optical phonon mode.

At $\nu =1.2,$ the two gapped branches are degenerate at $\mathbf{k}=0$. For 
$\nu >1.2,$ the spin-wave branch is connected to the phonon branch at point $%
M $ as in Fig. 4(a) while for $\nu \leq 1.2,$ it is connected to the $XY$
spin mode. The gap in both branches goes to zero as $\Delta _{Z}\rightarrow
0 $ so that, in the meron crystal, all four branches are gapless as seen in
Fig. 4(b).

We can understand the origin of the three gapless modes (in addition to the
phonon mode) in a spin-meron crystal when $\Delta_Z=0$ by the fact that the
energy of such a spin texture is invariant with respect to a rotation of the
spins around three orthogonal directions. Because these three symmetries are
spontaneously broken in the meron crystal, we get three Goldstone modes
corresponding, in our case, to oscillations in the $x-y,y-z$ and $x-z$
planes. This interpretation is confirmed by our animations of these modes%
\cite{animation}. There is no motion of the charge in any of these three
modes, the optical phonon mode present when $\Delta _{Z}\neq 0$, i.e. in the
Skyrme crystal, is presumably to be found amongst the higher-energy modes
not considered in our study. As $\Delta _{Z}\rightarrow 0$, the out of phase
motion of the Skyrmions in the optical phonon mode gradually stops until it
disappears completely in the meron crystal.

In Fig. 4 and other similar figures, we plot the poles with the two (six for
the $CP^{3}$ case below) biggest weights in the response functions indicated
in the legend. The function $\chi _{xx}$ stands for $\chi _{S_{x},S_{x}},$ $%
\chi _{yy}$ stands for $\chi _{S_{y},S_{y}}$ etc. For very small wavevector $%
\mathbf{k}$, this gives an indication of the nature of the mode. The phonon
mode, for example, is strongest in the density response $\chi _{n,n}$ while
the $XY$ mode is strongest in $\chi _{zz}$.

\begin{figure}[tbph]
\includegraphics[scale=1]{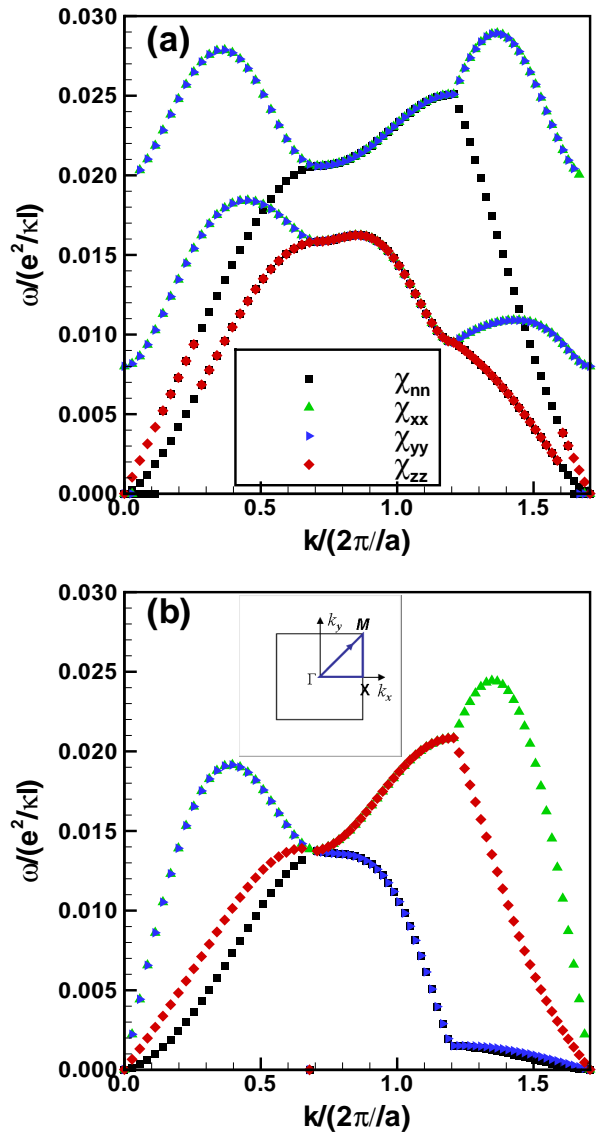}
\caption{(Color online) Dispersion relation of the collective modes of the
Skyrme crystal for a (a) SLA configuration with $\protect\nu =1.30$ and $%
\Delta _{Z}/\left( e^{2}/\protect\kappa \ell \right) =0.008$ and (b) a meron
configuration similar to that of Fig. 3(b) but with $\protect\nu =1.38$ and $%
\Delta _{Z}/\left( e^{2}/\protect\kappa \ell \right) =0.$ The dispersion is
computed along the path $\Gamma -M-X-\Gamma $ in the Brillouin zone as shown
by the inset in Fig. 4(b). The symbols indicate from what response function
each frequency comes. The softening of the modes seen in (b) is due to the
meron crystal becoming unstable at $\protect\nu =1.36$ with respect to a
biskyrmion crystal. }
\label{fig4a}
\end{figure}

\section{Pseudospin-Skyrmion crystals}

If the spin degree of freedom is neglected, then the 2DEG in a DQWS can be
described by its pseudospin and charge orders only. In the absence of bias,
the ground state at $\nu =1$ is a pseudospin ferromagnet\cite{macdobible}.
For nonzero interlayer separation $d,$ all pseudospins are forced to lie in
the $XY$ plane in order to minimise the capacitive energy (the term $%
J_{z,2}\left( 0\right) \left\vert \left\langle P_{z}\left( 0\right)
\right\rangle \right\vert ^{2}$ in Eq. (\ref{edouble}) below). The charged
excitations of this system are pseudospin-Skyrmions or bimerons\cite%
{macdobible}. A finite density of bimerons is included in the ground state
when $\nu \neq 1$. Once again, these topological excitations should form a
crystal at zero temperature. The Hartree-Fock ground-state energy is
written, in the pseudospin language, as%
\begin{eqnarray}
\frac{E_{HF}}{N} &=&-\frac{\Delta _{SAS}}{\nu }\left\langle P_{x}\left( 
\mathbf{0}\right) \right\rangle  \label{edouble} \\
&&+\frac{1}{4\nu }\sum_{\mathbf{G}}\Upsilon _{2}\left( \mathbf{G}\right)
\left\vert \left\langle \rho \left( \mathbf{G}\right) \right\rangle
\right\vert ^{2}  \notag  \label{edoublep} \\
&&+\frac{1}{\nu }\sum_{\mathbf{G}}J_{z,2}\left( \mathbf{G}\right) \left\vert
\left\langle P_{z}\left( \mathbf{G}\right) \right\rangle \right\vert ^{2} 
\notag \\
&&-\frac{1}{\nu }\sum_{\mathbf{G}}\widetilde{X}\left( \mathbf{G}\right)
\left\vert \left\langle \mathbf{P}_{\bot }\left( \mathbf{G}\right)
\right\rangle \right\vert ^{2},  \notag
\end{eqnarray}%
where 
\begin{equation}
J_{z,2}\left( \mathbf{G}\right) =H\left( \mathbf{G}\right) -\widetilde{H}%
\left( \mathbf{G}\right) -X\left( \mathbf{G}\right) ,
\end{equation}%
and 
\begin{equation}
\mathbf{P}_{\bot }\left( \mathbf{G}\right) =P_{x}\left( \mathbf{G}\right) 
\widehat{\mathbf{x}}+P_{y}\left( \mathbf{G}\right) \widehat{\mathbf{y}}.
\end{equation}

The interlayer coherence (i.e. $\left\langle \mathbf{P}_{\bot }\left( 
\mathbf{q}\right) \right\rangle \neq 0$) is lost at some critical interlayer
separation which is of the order of $d/\ell \approx 1.2$ at zero bias and
tunneling. In the coherent regime, the phase diagram in the $d/\ell -\Delta
_{SAS}-\nu $ space is very rich and has been only partially studied\cite%
{breycristalbimerons},\cite{cotegroup1}. In analogy with the phase diagram
of the Skyrme crystal in a SQWS, we find, in some regions of the parameter
space where $\Delta _{SAS}\neq 0,$ a solution consisting of a square lattice
with two bimerons of opposite global phases $\varphi $ and meron-antimeron
orientation reversed per unit cell. We call this phase an SLA bimeron
crystal. An example of such crystal is given in Fig. 5. Near zero tunneling,
we get a crystal of merons with four merons per unit cell in exactly the
same $SLA\ $configuration as that of the spin-meron crystal shown in Fig.
3(b).

\begin{figure}[tbph]
\includegraphics[scale=1]{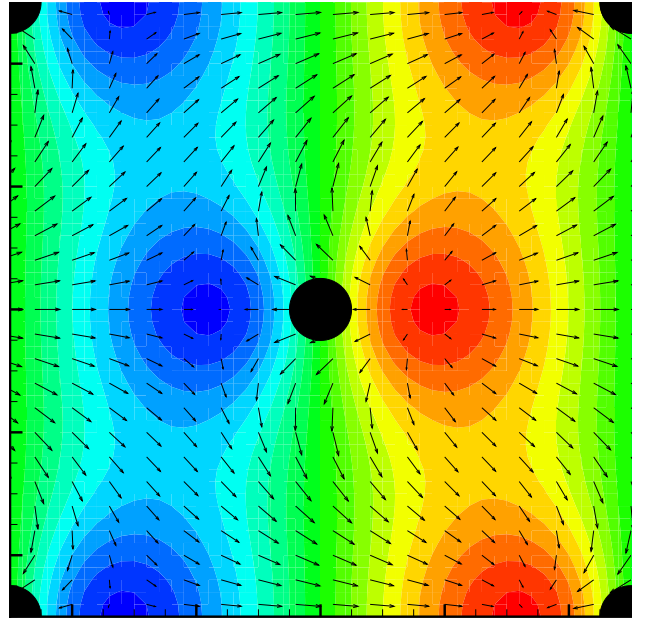}
\caption{(Color online) Pseudospin texture in the $x-y$ plane (vectors) and
value of $S_{z}$ (contours) in a unit cell of a bimeron crystal in a SLA
configuration with $\protect\nu =1.20$ and $\Delta _{SAS}/\left( e^{2}/%
\protect\kappa \ell \right) =0.02.$ The full black circles indicate the
position of the maximum in the bimeron density. Dark (red) contours
correspond to $S_{z}=+\frac{1}{2}$ while gray (blue) contours are for $%
S_{z}=-\frac{1}{2}.$ Note that the pseudospin texture for the bimeron at the
center has a global phase rotation of with respect to the pseudospin texture
of the bimerons at the four corners of the unit cell. }
\end{figure}

The collective mode spectrum of a bimeron crystal is shown in Fig. 6 where
again we have plotted the four lowest branches only. There is one gapless
and three gapped modes. The gapless mode is the phonon mode, which appears
as a pole of $\chi _{nn}$ and has a $\omega \sim k^{3/2}$ dispersion as in
the Skyrme crystal. The next lowest mode has a very small gap that closes as
the interlayer separation $d\rightarrow 0$ \textit{even when tunneling is
finite}. In this gapless limit it has a linear dispersion. This behavior is
easily understood: when $d=0$ and $\Delta _{SAS}=0$, a bimeron crystal can
be mapped onto a Skyrme crystal with zero Zeeman coupling. (A bimeron is
obtained by rotating the spin texture of a Skyrmion by $90$ degrees around
the spin $x$ or $y$ axis.) If one rotates around the $y$ axis, the resulting
bimeron has pseudospins lying along the $x$ direction at infinity, with the
two merons aligned along the $x$ axis. The tunneling term $\Delta
_{SAS}S_{x} $ for the bimeron crystal then plays the role of the Zeeman term 
$\Delta _{Z}S_{z}$ in the Skyrme crystal. It follows that, for $d=0$, the
collective modes of the bimeron crystal are identical to that of Skyrme
crystal with $\Delta _{Z}=\Delta _{SAS}.$ However, the specific modes appear
in different response functions. When $d=0$ but tunneling is finite, Eq. (%
\ref{edouble}) becomes

\begin{eqnarray}
\frac{E_{HF}}{N} &=&-\frac{\Delta _{SAS}}{\nu }\left\langle P_{x}\left( 
\mathbf{0}\right) \right\rangle  \label{pseudo} \\
&&+\frac{1}{4\nu }\sum_{\mathbf{G}}\Upsilon _{2}\left( \mathbf{G}\right)
\left\vert \left\langle \rho \left( \mathbf{G}\right) \right\rangle
\right\vert ^{2}  \notag \\
&&-\frac{1}{\nu }\sum_{\mathbf{G}}X\left( \mathbf{G}\right) \left\vert
\left\langle \mathbf{P}\left( \mathbf{G}\right) \right\rangle \right\vert
^{2},  \notag
\end{eqnarray}%
From Eq. (\ref{pseudo}), we see that the pseudospins can rotate freely in
the $y-z$ plane but not in the $x-y$ plane if $\Delta _{SAS}\neq 0$. The $XY$
Goldstone mode of the Skyrme crystal becomes a $YZ$ Goldstone mode in the
bimeron crystal at $d=0$. At finite interlayer separation $H\left( \mathbf{G}%
\right) \neq \widetilde{H}\left( \mathbf{G}\right) $ and $X\left( \mathbf{G}%
\right) \neq \widetilde{X}\left( \mathbf{G}\right) $ and a rotation in the $%
y-z$ plane can no longer be performed without energy cost. The $YZ$ mode is
thus gapped at finite interlayer separation.

We can explain the smallnest of this gap in the following way. A global
rotation of the pseudospins in the $y-z$ plane leads to a solid rotation of
the bimerons (i.e. a rotation of the pairs of merons around the filled black
circles in Fig. 5). At $d=0$, the bimeron density is circularly symmetric%
\cite{breycristalbimerons} and so does not change under a rotation of the
bimeron. At finite $d$, however, the bimeron density is no longer circularly
symmetric. A rotation of an isolated bimeron would not change its energy
but, in a crystal, these charged objects interact with one another via the
Coulomb interaction, so there is an energy cost for such rotations. The
energy cost, however, is small because it comes from high order multipole
interactions. (The rotation of the bimerons can be clearly seen in the
animations\cite{animation}.)

At $d\neq 0$ but $\Delta _{SAS}=0$ (in which case we get a meron crystal), a
rotation of the pseudospins in the $x-y$ plane without an energy cost is
possible. This in this limit a meron crystal has a gapless $XY\ $mode (see
Fig. 7) just as a Skyrme crystal.

\begin{figure}[tbph]
\includegraphics[scale=1]{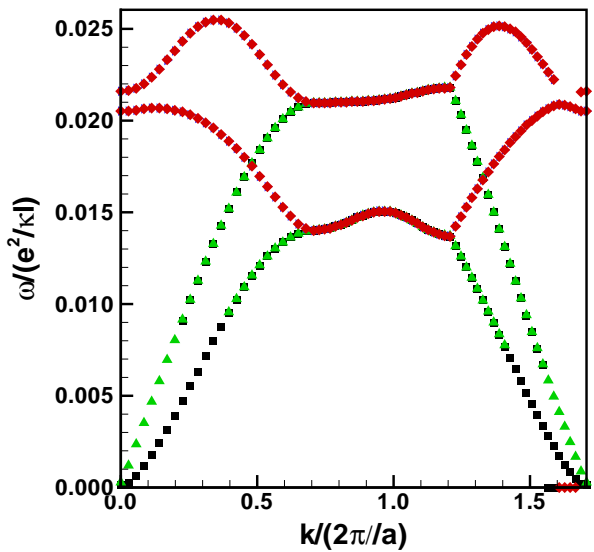}
\caption{(Color online) Collective mode spectrum of a bimeron crystal in the
SLA configuration of Fig. 5 with $\protect\nu =1.2,\Delta _{SAS}/\left(
e^{2}/\protect\kappa \ell \right) =0.02$ and $d/\ell =0.1.$ The path in the
Brillouin zone is as indicated in the inset of Fig. 4(b) and the symbols are
as indicated in the legend of that same figure. }
\label{fig4}
\end{figure}

The interlayer-coherent homogeneous phase at $\nu =1$ can sustain\cite%
{fertigdispersion} a pseudospin-wave mode in which the pseudospins execute
small precession around their local equilibrium direction (the $x$ axis for $%
\Delta _{SAS}\neq 0$). The pseudospin-wave mode is equivalent to the
spin-wave (or Zeeman)\ mode in the Skyrme crystal. We discussed this mode
before in Sec. IV. Its dispersion is given by Eq. (\ref{pseudowave}). For $%
\Delta _{SAS}=0$, the dispersion $\omega _{5}\left( q\right) $ is linear in $%
q$ for small wavevector. This mode has been detected experimentally\cite%
{spielmanpseudospinmode}. For $\Delta _{SAS}\neq 0$ and $d=0$, the
pseudospin-wave mode is gapped at $\omega _{PSW}\left( 0\right) =\Delta
_{SAS}$ as can be seen from Eq. (\ref{pseudowave}). At finite $d$, that gap
is increased. In Fig. 6, $d$ is small and we see two modes gapped at a value
near $\Delta _{SAS}$. If we replace $\left\langle \rho _{1,3}\left( 0\right)
\right\rangle ,\nu _{R}$ and $\nu _{L}$ by their values in the crystal
phases in Eq. (\ref{pseudowave}), we get a very good estimate of the
pseudospin-wave gap $\omega _{5}\left( 0\right) $ in the bimeron crystal. In
the example of Fig. 6, $d/\ell =0.1,$ $\left\langle \rho _{1,3}\left(
0\right) \right\rangle =0.106$ and $\nu _{R}=\nu _{L}=0.6$ so that we find $%
\omega _{5}\left( 0\right) =0.021\left( e^{2}/\hslash \kappa \ell \right) $
which is in good agreement with the computed gap. We find excellent
agreement for other values of $d/\ell $ as well. This permits us to identify
the higher-energy dispersion branch as the pseudospin-wave mode. Our
animations\cite{animation} show that the other high-energy branch is an
optical phonon mode in which the two bimerons in the unit cell have opposite
motions.

\begin{figure}[tbph]
\includegraphics[scale=1]{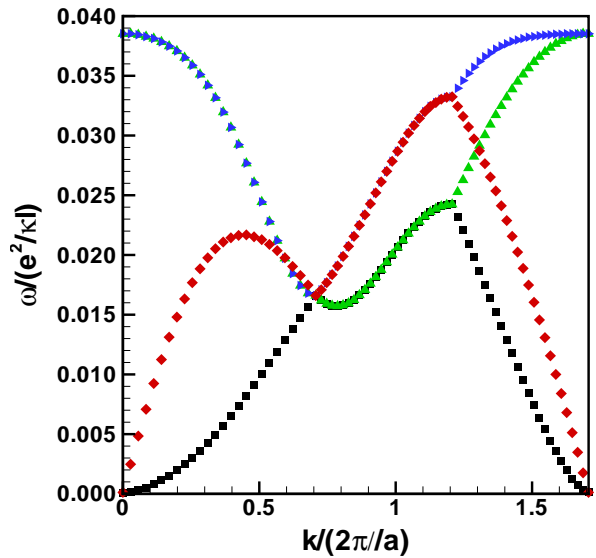}
\caption{(Color online) Collective mode spectrum of a pseudospin meron
crystal in a SLA configuration similar to that of Fig. 3(b) with $\protect%
\nu =1.2,\Delta _{SAS}/\left( e^{2}/\protect\kappa \ell \right) =0$ and $%
d/\ell =0.8$. The path in the Brillouin zone is as indicated in the inset of
Fig. 4(b) and the symbols are as indicated in the legend of that figure. }
\end{figure}

Figure 7 shows the collective mode spectrum of a pseudospin meron crystal.
The two gapless modes are the phonon and the $XY$ mode as discussed above
and the other two branches are the optical phonon and pseudospin-wave modes.
These two branches are degenerate along $\Gamma M$ at all $d/\ell $ for the
SLA\ configuration. Eq. (\ref{pseudowave}) fails in this case since $%
\left\langle \rho _{1,3}\left( 0\right) \right\rangle =0$ for a meron
crystal (but $\left\langle \rho _{1,3}\left( \mathbf{G}\neq 0\right)
\right\rangle \neq 0$) and we get $\omega _{5}\left( 0\right) =0$ for all
interlayer separations. That this equation works for the bimeron crystal but
fails for the meron crystal can be understood by the fact that the bimeron
crystal has regions of coherent liquid between the bimerons whereas the
meron crystal has no such regions.

\section{CP$^{3}$ Skyrmion crystals}

Now that we understand the limiting cases of pure spin and pure pseudospin
crystals, we are ready to look at the collective excitations of the $CP^{3}$
Skyrmion crystal with has entangled spin and pseudospin textures. We study
the excitation spectrum as a function of interlayer separation, tunneling,
and bias as was done for the ground state in Paper I\cite{cotecp3}. We refer
the reader to this paper for the definition of the various groundstates that
we will use in the present section. We remark that all our data for the $%
CP^{3}$ Skyrmion crystal in this section are for filling factor $\nu =0.8$
where we get a good convergence of the order parameters for the $CP^{3}$
phases. For $\nu >1.0,$ a similar phase is difficult to find numerically
(see Paper I\cite{cotecp3}). For the spin-Skyrmion and pseudospin-Skyrmion
crystals, our results above were for $\nu >1.0.$ In these two cases,
however, the phase diagram below $\nu =1$, can be related to that above $\nu
=1$ by particle-hole symmetry. In the $CP^{3}$ Skyrmion case, there is no
such particle-hole symmetry.

The groundstate energy of Eq. (\ref{gsenergy}) can be written in the
spin-pseudospin langage as%
\begin{gather}
\frac{E_{HF}}{N}=\frac{\Delta _{b}}{\nu }\left\langle P_{z}\left( \mathbf{0}%
\right) \right\rangle -\frac{\Delta _{Z}}{\nu }\left\langle S_{z}\left( 
\mathbf{0}\right) \right\rangle -\frac{\Delta _{SAS}}{\nu }\left\langle
P_{x}\left( \mathbf{0}\right) \right\rangle  \notag \\
+\frac{1}{4\nu }\sum_{\mathbf{G}}\Upsilon _{1}\left( \mathbf{G}\right)
\left\vert \left\langle \rho \left( \mathbf{G}\right) \right\rangle
\right\vert ^{2}+\frac{1}{\nu }\sum_{\mathbf{G}}J_{z,1}\left( \mathbf{G}%
\right) \left\vert \left\langle P_{z}\left( \mathbf{G}\right) \right\rangle
\right\vert ^{2}  \notag \\
-\frac{1}{\nu }\sum_{\mathbf{G}}\sum_{a=R,L}X\left( \mathbf{G}\right)
\left\vert \left\langle \mathbf{S}_{a}\left( \mathbf{G}\right) \right\rangle
\right\vert ^{2}  \label{enerpseudos} \\
-\frac{1}{\nu }\sum_{\mathbf{G}}\sum_{\alpha =+,-}\widetilde{X}\left( 
\mathbf{G}\right) \left[ \left\vert \left\langle P_{x,\alpha }\left( \mathbf{%
G}\right) \right\rangle \right\vert ^{2}+\left\vert \left\langle P_{y,\alpha
}\left( \mathbf{G}\right) \right\rangle \right\vert ^{2}\right]  \notag \\
-\frac{1}{\nu }\sum_{\mathbf{G}}\widetilde{X}\left( \mathbf{G}\right) \left[
\left\vert \left\langle \rho _{1,4}\left( \mathbf{G}\right) \right\rangle
\right\vert ^{2}+\left\vert \left\langle \rho _{2,3}\left( \mathbf{G}\right)
\right\rangle \right\vert ^{2}\right] ,  \notag
\end{gather}%
where we have defined the interactions 
\begin{eqnarray}
\Upsilon _{1}\left( \mathbf{G}\right) &=&H\left( \mathbf{G}\right) +%
\widetilde{H}\left( \mathbf{G}\right) -\frac{1}{2}X\left( \mathbf{G}\right) ,
\\
J_{z,1}\left( \mathbf{G}\right) &=&H\left( \mathbf{G}\right) -\widetilde{H}%
\left( \mathbf{G}\right) -\frac{1}{2}X\left( \mathbf{G}\right) ,
\end{eqnarray}%
and the fields (see Eq. (\ref{a3}))%
\begin{equation}
P_{x,+}\left( \mathbf{G}\right) =\frac{1}{2}\left[ \left\langle \rho
_{1,3}\right\rangle +\left\langle \rho _{3,1}\right\rangle \right] ,
\end{equation}%
\begin{equation}
P_{x,-}\left( \mathbf{G}\right) =\frac{1}{2}\left[ \left\langle \rho
_{2,4}\right\rangle +\left\langle \rho _{4,2}\right\rangle \right] ,
\end{equation}%
\begin{equation}
P_{y,+}\left( \mathbf{G}\right) =\frac{1}{2i}\left[ \left\langle \rho
_{1,3}\right\rangle -\left\langle \rho _{3,1}\right\rangle \right] ,
\end{equation}%
\begin{equation}
P_{y,-}\left( \mathbf{G}\right) =\frac{1}{2i}\left[ \left\langle \rho
_{2,4}\right\rangle -\left\langle \rho _{4,2}\right\rangle \right] .
\end{equation}

In Eq. (\ref{enerpseudos}), $S_{i}=\sum_{a=R,L}S_{i,a}$ and $%
P_{i}=\sum_{\alpha =+,-}P_{i,\alpha }$ where $i=x,y,z.$ The total density is
defined by 
\begin{equation}
\left\langle \rho \left( \mathbf{G}\right) \right\rangle =\left\langle \rho
_{11}\left( \mathbf{G}\right) \right\rangle +\left\langle \rho _{22}\left( 
\mathbf{G}\right) \right\rangle +\left\langle \rho _{33}\left( \mathbf{G}%
\right) \right\rangle +\left\langle \rho _{44}\left( \mathbf{G}\right)
\right\rangle .
\end{equation}%
We will also make used of the fields%
\begin{eqnarray}
R_{xx}(\mathbf{q}) &=&\frac{1}{2}\left[ \rho _{1,4}(\mathbf{q})+\rho _{4,1}(%
\mathbf{q})+\rho _{2,3}(\mathbf{q})+\rho _{3,2}(\mathbf{q})\right] ,
\label{rxxx} \\
R_{zz}(\mathbf{q}) &=&\frac{1}{2}\left[ \rho _{1,1}(\mathbf{q})-\rho _{2,2}(%
\mathbf{q})-\rho _{3,3}(\mathbf{q})+\rho _{4,4}(\mathbf{q})\right] .
\end{eqnarray}

\subsection{Variation with interlayer separation}

At small value $\Delta _{SAS}/\left( e^{2}/\kappa \ell \right)
=0.0002<<\Delta _{Z}/\left( e^{2}/\kappa \ell \right) $ of the tunneling
parameter, the ground state of the 2DEG at small interlayer separation $%
d/\ell $ is a spin-polarized meron crystal with an SLA\ configuration\cite%
{cotecp3}. The corresponding excitation spectrum is shown in Fig. 8. For all
the following dispersion plots, we represent the poles of $\chi _{nn}$ by
the full (black) circles, those of the pseudospin response functions $\chi _{%
\mathbf{PP}}$ by the empty (blue) squares, those of the spin response
functions $\chi _{\mathbf{SS}}$ by the right (red) triangles and by the left
(green)\ triangles the poles of the two response functions $\chi
_{R_{xx},R_{xx}}$ and $\chi _{R_{zz},R_{zz}}$. The operator $R_{xx}$ (see
Eq. (\ref{rxxx})) involves a change in both the spin and pseudospin indices
unlike the operators $\mathbf{S},\mathbf{P}$ or $n$. There are clearly two
basic mode structures in Fig. 8. One, noted \textquotedblleft
(1)\textquotedblright\ is the pseudospin meron crystal response already
studied in Fig. 7 and it comprises four branches. The second structure,
noted \textquotedblleft (2)\textquotedblright , has two twice-degenerate
branches for a total of eight branches for the whole $CP^{3}$ crystal
(considering the low-energy modes only since other higher-energy modes are
also present in the GRPA). One branch in (2) is the Zeeman mode (gapped at $%
\Delta _{Z}/\left( e^{2}/\kappa \ell \right) =0.004$) while the other, which
is also gapped, seems to appear predominantly in $\chi _{R_{xx}R_{xx}}$ and $%
\chi _{R_{zz}R_{zz}}$.

\begin{figure}[tbph]
\includegraphics[scale=1]{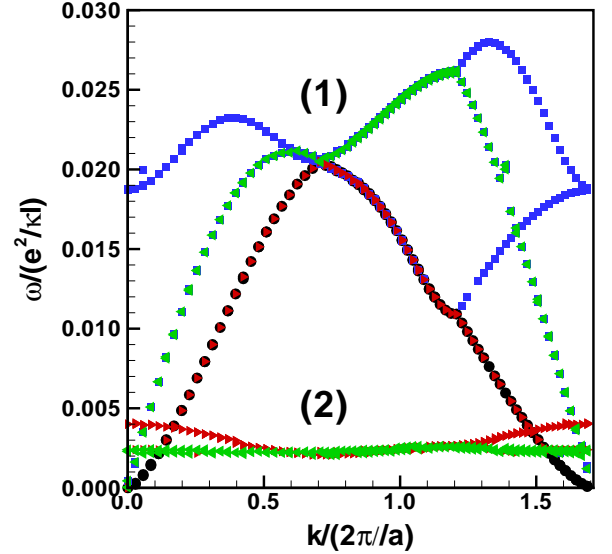}
\caption{(Color online) Collective mode spectrum of a spin-polarized meron
crystal with parameters: $\protect\nu =0.8,\Delta _{Z}/\left( e^{2}/\protect%
\kappa \ell \right) =0.004,\Delta _{SAS}/\left( e^{2}/\protect\kappa \ell
\right) =0.0002,\Delta _{b}/\left( e^{2}/\protect\kappa \ell \right) =0,$
and $d/\ell =0.3.$ }
\end{figure}

Figure 9 shows what happens to this collective mode spectrum\ when we
increase the interlayer separation and keep both $\Delta _{SAS}$ and $\Delta
_{Z}$ fixed. In this case (see Fig. 3 of Paper I\cite{cotecp3}), the ground
state becomes a $CP^{3}$ crystal at a critical layer separation $d_{c}$ that
increases with $\Delta _{Z}$. For $\Delta _{Z}/\left( e^{2}/\kappa \ell
\right) =0.004$, we find $d_{c}/\ell \approx 0.6$. In Fig. 9, $d/\ell =1.0$
and we see that the degeneracy of the modes in the structure (2) of Fig. 6
has been lifted and that \textit{two} seemingly gapless modes have emerged
from the lowest branch. In fact, the gap in the lowest branch in structure
(2) vanishes continuously as the transition is approached. We believe that
this should be observable experimentally. In the following analysis, it will
become clear that of these two modes, one only one is truly gapless. We have
identified the phonon, pseudospin-wave, pseudospin-$YZ$, and Zeeman modes in
Fig. 9. We find that at $d/\ell =1.1$, the modes of structure (2) become
soft at point $M$ in the Brillouin zone indicating an instability of the $%
CP^{3}$ crystal analogous to that of the uniform ($\nu =1$) system at
similar separations \cite{fertigdispersion}.

\begin{figure}[tbph]
\includegraphics[scale=1]{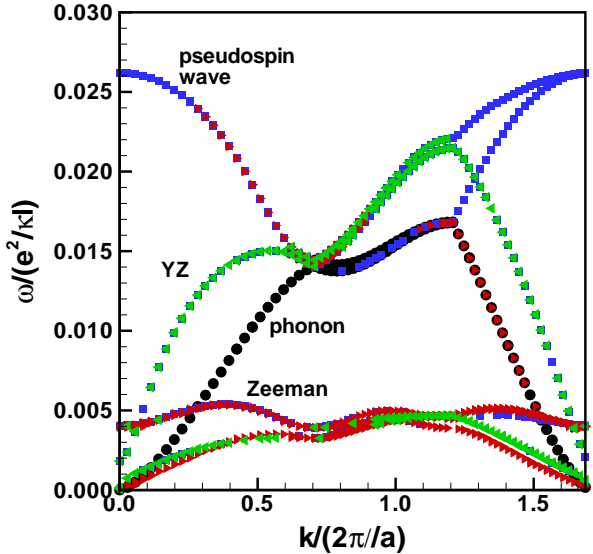}
\caption{(Color online) Collective mode spectrum of a crystal of $CP^{3}$
Skyrmions along path -M-X- in the Brillouin zone. The parameters are: $%
\protect\nu =0.8,\Delta _{Z}/\left( e^{2}/\protect\kappa \ell \right)
=0.004,\Delta _{SAS}/\left( e^{2}/\protect\kappa \ell \right) =0.0002,\Delta
_{b}/\left( e^{2}/\protect\kappa \ell \right) =0,$ and $d/\ell =1.0.$}
\end{figure}

\subsection{Variation with tunnel coupling}

To check if the two modes in Fig. 9 are really gapless, we plot in Fig. 10
the collective mode spectrum of the $CP^{3}$ crystal at a higher value of
tunneling. In Paper I\cite{cotecp3} it was shown that when tunneling is
increased, there is a transition from the $CP^{3}$ crystal state to a
symmetric Skyrmion (SS) state. The SS is basically a pseudospin polarized
spin-Skyrmion state in which the $S,\uparrow $ and $S,\downarrow $ only are
occupied. There is thus a spin texture on top of a state where all
pseudospins point along the $x$ axis. For Fig. 10, we choose $d/\ell
=1.0,\Delta _{Z}/\left( e^{2}/\kappa \ell \right) =0.008$ and $\Delta
_{SAS}/\left( e^{2}/\kappa \ell \right) =0.0044$. From Fig. 5(c) of Paper I,
the ground state of the 2DEG is a SS state in this case. The $CP^{3}$
Skyrmion state is still stable even if is not the ground state and, as we
see from Fig. 10, there is a very small gap with energy smaller than $%
0.001\left( e^{2}/\kappa \ell \right) $ in one of the two modes of structure
2. As we will show below, the application of an electrical bias does not
open new gaps in the system so that we conclude that with finite bias,
Zeeman, and tunneling energies, our $CP^{3}$ crystal has \textit{two}
Goldstone modes: the phonon mode and a phase mode whose origin we will
discuss below.

\begin{figure}[tbph]
\includegraphics[scale=1]{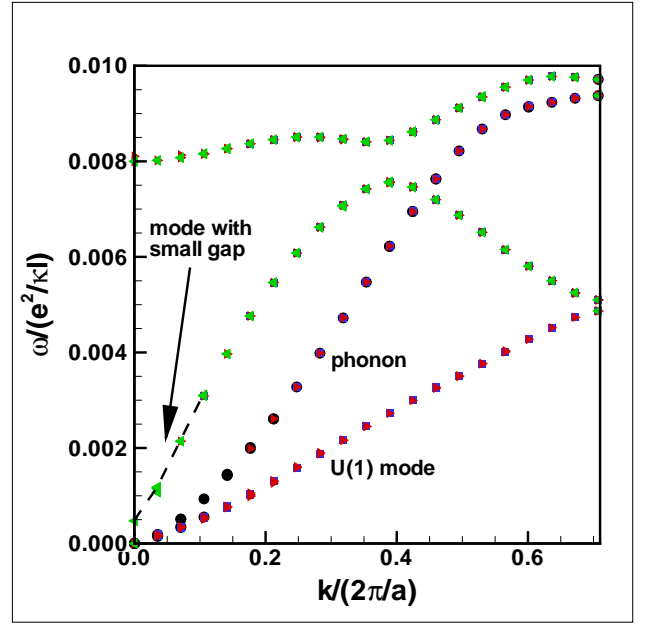}
\caption{(Color online) Collective mode spectrum of a crystal of $CP^{3}$
Skyrmions with strong tunneling. The parameters are: $\protect\nu %
=0.8,\Delta _{Z}/\left( e^{2}/\protect\kappa \ell \right) =0.008,\Delta
_{SAS}/\left( e^{2}/\protect\kappa \ell \right) =0.0044,\Delta _{b}/\left(
e^{2}/\protect\kappa \ell \right) =0,$ and $d/\ell =1.0.$}
\end{figure}

The collective mode spectrum of a SS state at $d/\ell =1.0,\Delta
_{Z}/\left( e^{2}/\kappa \ell \right) =0.008$ and $\Delta _{SAS}/\left(
e^{2}/\kappa \ell \right) =0.0044$ is shown in Fig. 11. For $\Delta
_{Z}/\left( e^{2}/\kappa \ell \right) =0.008$, the transition to the SS
state occurs at $\Delta _{SAS}^{(c)}/\left( e^{2}/\kappa \ell \right)
\approx 0.004$ which is roughly $\Delta _{SAS}^{(c)}\approx 0.5\Delta _{Z}$.
The phonon and spin mode dispersions in Fig. 11 modes are typical of that of
a Skyrme crystal (see Fig. 4(a)). There are only two gapless modes: the
phonon and spin $XY$ phase mode.

\begin{figure}[tbph]
\includegraphics[scale=1]{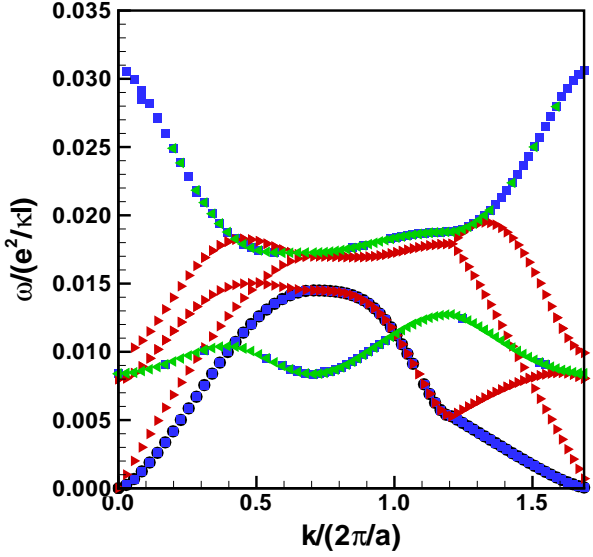}
\caption{(Color online) Collective mode spectrum of a crystal of symmetric
Skyrmions. The parameters are: $\protect\nu =0.8,\Delta _{Z}/\left( e^{2}/%
\protect\kappa \ell \right) =0.008,\Delta _{Z}/\left( e^{2}/\protect\kappa %
\ell \right) =0.006,\Delta _{b}/\left( e^{2}/\protect\kappa \ell \right) =0$%
, and $d/\ell =1.0.$}
\end{figure}

\subsection{Variation with electrical bias}

We now turn on the bias between the two wells so that all the charge is
gradually transferred to the left well. If we start with a $CP^{3}$ crystal
at zero bias, our Hartree-Fock analysis (see Fig. 9 of Paper I\cite{cotecp3}%
) shows that the spin and pseudospin textures are maintained until the
charge has completely gone into one well and we get a spin-Skyrmion crystal.
From the collective mode spectrum in Fig. 12, we see that the two Goldstone
modes are preserved at finite bias if we consider that one of the low-energy
mode has an extremely small gap (whose origin we explain in the next
section). The qualitative features of the spectrum are unchanged from that
of Fig. 9. The $CP^{3}$ crystal is stable up to the largest bias for which
there is a charge in the right well. When the charge is completely
transferred to the left well, the spectrum is that of the Skyrme crystal
shown in Fig. 4(a).

\begin{figure}[tbph]
\includegraphics[scale=1]{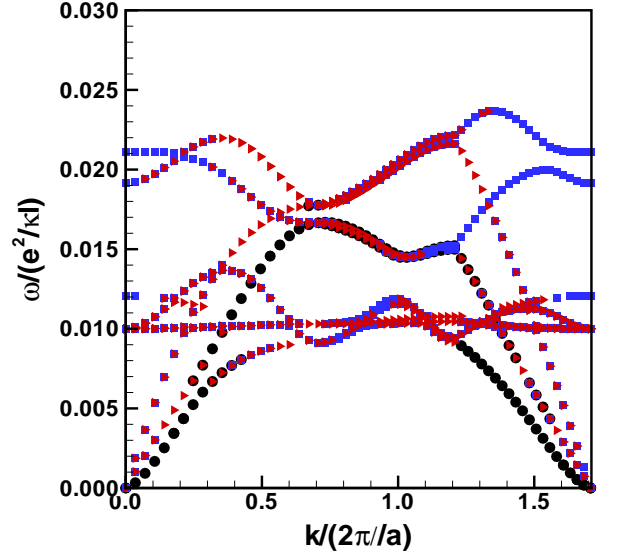}
\caption{(Color online) Collective mode spectrum of a crystal of $CP^{3}$
Skyrmions with a finite bias. The parameters are: $\protect\nu =0.8,\Delta
_{Z}/\left( e^{2}/\protect\kappa \ell \right) =0.01,\Delta _{SAS}/\left(
e^{2}/\protect\kappa \ell \right) =0.0002,\Delta _{b}/\left( e^{2}/\protect%
\kappa \ell \right) =0.20$ and $d/\ell =1.0.$}
\end{figure}

\subsection{Physical origin of the Goldstone modes}

Following Gosh and Rajaraman\cite{rajaramancp3}, we pametrize a general $%
CP^{3}$ spinor by six angles that can be roughly interpreted as the polar
and azimuthal angles of the spins in the right and left wells: $\theta
_{R},\theta _{L}$ and $\varphi _{R},\varphi _{L},$ and the polar and
azimuthal angles of the total pseudospin, $\alpha $ and $\beta $. Writing%
\begin{equation}
a\left( X\right) =\left( 
\begin{array}{c}
\cos \frac{\alpha }{2}\cos \frac{\theta _{R}}{2} \\ 
\cos \frac{\alpha }{2}\sin \frac{\theta _{R}}{2}e^{i\varphi _{R}} \\ 
\sin \frac{\alpha }{2}\cos \frac{\theta _{L}}{2}e^{i\beta } \\ 
\sin \frac{\alpha }{2}\sin \frac{\theta _{L}}{2}e^{i\left( \beta +\varphi
_{L}\right) }%
\end{array}%
\right) ,  \label{23_3}
\end{equation}%
a possible {spin-pseudospin texture state can be written as%
\begin{equation}
\left\vert \Psi \right\rangle =\prod\limits_{X}\left[ \sum_{\sigma
}a_{\sigma }\left( X\right) c_{\sigma ,X}^{\dag }\right] \left\vert
0\right\rangle ,  \label{23_2}
\end{equation}%
where $\left\vert 0\right\rangle $ is the vacuum state. We replace\cite%
{rajaramancp3} $X$ by $\mathbf{r}$ in this last equation so that our order
parameters can be written in real space as }

\begin{equation}
\left\langle \rho _{\sigma _{1},\sigma _{2}}\left( \mathbf{r}\right)
\right\rangle =a_{\sigma _{1}}^{\ast }\left( \mathbf{r}\right) a_{\sigma
_{2}}\left( \mathbf{r}\right) .
\end{equation}%
We have

\begin{eqnarray}
\left\langle \rho _{1,1}\left( \mathbf{r}\right) \right\rangle &=&\cos ^{2}%
\frac{\alpha \left( \mathbf{r}\right) }{2}\cos ^{2}\frac{\theta _{R}\left( 
\mathbf{r}\right) }{2}, \\
\left\langle \rho _{2,2}\left( \mathbf{r}\right) \right\rangle &=&\cos ^{2}%
\frac{\alpha \left( \mathbf{r}\right) }{2}\sin ^{2}\frac{\theta _{R}\left( 
\mathbf{r}\right) }{2},  \notag \\
\left\langle \rho _{3,3}\left( \mathbf{r}\right) \right\rangle &=&\sin ^{2}%
\frac{\alpha \left( \mathbf{r}\right) }{2}\cos ^{2}\frac{\theta _{L}\left( 
\mathbf{r}\right) }{2},  \notag \\
\left\langle \rho _{4,4}\left( \mathbf{r}\right) \right\rangle &=&\sin ^{2}%
\frac{\alpha \left( \mathbf{r}\right) }{2}\sin ^{2}\frac{\theta _{L}\left( 
\mathbf{r}\right) }{2},  \notag
\end{eqnarray}%
and%
\begin{eqnarray}
\left\langle \rho _{1,2}\left( \mathbf{r}\right) \right\rangle &=&\frac{1}{2}%
\left( \cos ^{2}\frac{\alpha }{2}\sin \theta _{R}\right) e^{i\varphi _{R}},
\\
\left\langle \rho _{1,3}\left( \mathbf{r}\right) \right\rangle &=&\frac{1}{2}%
\left( \sin \alpha \cos \frac{\theta _{L}}{2}\cos \frac{\theta _{R}}{2}%
\right) e^{i\beta },  \notag \\
\left\langle \rho _{1,4}\left( \mathbf{r}\right) \right\rangle &=&\frac{1}{2}%
\left( \sin \alpha \sin \frac{\theta _{L}}{2}\cos \frac{\theta _{R}}{2}%
\right) e^{i\beta +i\varphi _{L}},  \notag \\
\left\langle \rho _{2,3}\left( \mathbf{r}\right) \right\rangle &=&\frac{1}{2}%
\left( \sin \alpha \cos \frac{\theta _{L}}{2}\sin \frac{\theta _{R}}{2}%
\right) e^{i\beta -i\varphi _{R}},  \notag \\
\left\langle \rho _{2,4}\left( \mathbf{r}\right) \right\rangle &=&\frac{1}{2}%
\left( \sin \alpha \sin \frac{\theta _{L}}{2}\sin \frac{\theta _{R}}{2}%
\right) e^{-i\varphi _{R}}e^{i\beta +i\varphi _{L}},  \notag \\
\left\langle \rho _{3,4}\left( \mathbf{r}\right) \right\rangle &=&\frac{1}{2}%
\left( \sin ^{2}\frac{\alpha }{2}\sin \theta _{L}\right) e^{i\varphi _{L}}, 
\notag
\end{eqnarray}%
and all angles are functions of $\mathbf{r.}$

The symmetry-breaking fields in our Hamiltonian in Eq. (\ref{gsenergy})
couple to $P_{x},$ $P_{z},$ and $S_{z}$ where

\begin{align}
\left\langle S_{z}\left( \mathbf{r}\right) \right\rangle & =\frac{1}{2}\left[
\sin ^{2}\frac{\alpha }{2}\cos \theta _{L}+\cos ^{2}\frac{\alpha }{2}\cos
\theta _{R}\right] ,  \label{px} \\
\left\langle P_{x}\left( \mathbf{r}\right) \right\rangle & =\frac{1}{2}\sin
\alpha \left[ \cos \frac{\theta _{L}}{2}\cos \frac{\theta _{R}}{2}\cos \beta %
\right]  \notag \\
& +\frac{1}{2}\sin \alpha \left[ \sin \frac{\theta _{L}}{2}\sin \frac{\theta
_{R}}{2}\cos \left( \beta +\varphi _{L}-\varphi _{R}\right) \right] ,  \notag
\\
\left\langle P_{z}\left( \mathbf{r}\right) \right\rangle & =\frac{1}{2}\cos
\alpha .  \notag
\end{align}%
These expressions may be substituted into our Hartree-Fock Hamiltonian of
Eq. (\ref{gsenergy}). The origin of the second Goldstone mode (the phase
mode)\ is then clear: the energy of the $CP^{3}$ crystal is unchanged if we
change the azimuthal angles $\varphi _{R}$ and $\varphi _{L}$ by the same
amount (i.e., keep $\varphi _{R}-\varphi _{L}$ fixed) for all spins,
whereas, in the $CP^{3}$ structure, the symmetry in $\varphi _{R}+\varphi
_{L}$ is broken. The second Goldstone mode of the $CP^{3}$ crystal is thus
an $XY$ phase mode corresponding to a global rotation of the spins at $%
\mathbf{q}=0$ i.e. an $XY$ phase mode in $\varphi _{R}+\varphi _{L}.$

Were it not for the term $\cos \left( \beta +\varphi _{L}-\varphi
_{R}\right) $ in the expression of $P_{x}$ in Eq. (\ref{px}), our
Hamiltonian would also be invariant with respect to $\varphi _{R}-\varphi
_{L}$ and to $\beta $: we could then have two additional Goldstone modes.
The tunnel coupling is of the form $\Delta _{SAS}\int d\mathbf{r}%
\left\langle P_{x}\left( \mathbf{r}\right) \right\rangle $. In our numerical
solutions we find that the $CP^{3}$ crystal phase is stable at very small
values of $\Delta _{SAS}$ and, furthermore, at these values $\left\langle
P_{x}\left( \mathbf{r}\right) \right\rangle $ is also small. For example, in
the case of Fig. 12, we have $\int d\mathbf{r}\left\langle P_{x}\left( 
\mathbf{r}\right) \right\rangle =0.005$ and $\Delta _{SAS}/\left(
e^{2}/\kappa \ell \right) =0.0002$ so that a change in $\varphi _{R}-\varphi
_{L}$ or $\beta $ would lead to a change in the energy (a bare gap) $\Delta
_{SAS}\int d\mathbf{r}\left\langle P_{x}\left( \mathbf{r}\right)
\right\rangle $ by a quantity smaller than $1\times 10^{-6}\left(
e^{2}/\kappa \ell \right) $ which is smaller than our numerical accuracy.
Even if we allow for this change in energy to be renormalized by vertex
corrections (as is the case for the pseudospin-wave mode in the UCS in Sec.
IV where the gap $\Delta _{SAS}^{\ast }\approx 10\Delta _{SAS}$), the
expected gap would still be small. This is the reason why there seems to be
three gapless modes in Fig. 9. In Fig. 10, we have choosen $\Delta
_{SAS}/\left( e^{2}/\kappa \ell \right) =0.0044$ and obtain $\int d\mathbf{r}%
\left\langle P_{x}\left( \mathbf{r}\right) \right\rangle =0.163$ in this
case. We could thus expect a bare gap of order $0.0007\left( e^{2}/\kappa
\ell \right) $ which is consistent with the gap we find numerically. We thus
associate the small gap mode with either $\varphi _{R}-\varphi _{L}$ or $%
\beta $ as these angles are present in the expression of both $\mathbf{S}%
_{\bot }$ and $\mathbf{P}_{\bot }$. Presumably one of these angles (or a
combination of them) leads to the mode with the small gap

One last point is to check that a change $\varphi _{R},\varphi _{L}$ or $%
\beta $ does not change the Skyrmion density. We recall that the topological
charge density of $CP^{3}$ skyrmion is defined\cite{rajaramancp3} by

\begin{equation}
q\left( \mathbf{r}\right) =-\frac{i}{2\pi }\varepsilon _{\mu \nu }\left(
D_{\mu }a_{\sigma }\right) ^{\ast }\left( D_{\nu }a_{\sigma }\right) ,
\end{equation}%
where $D_{\mu }$ is the covariant derivative of the $U(1)$ gauge
transformation with the gauge defined by $A_{\mu }\left( \mathbf{r}\right)
=i\sum_{\sigma }a_{\sigma }^{\ast }\partial _{\mu }a_{\sigma }$ i.e. $D_{\mu
}=\partial _{\mu }+iA_{\mu }.$ With the parametrization given by Eq. (\ref%
{23_3}), we have 
\begin{equation}
q\left( \mathbf{r}\right) =\frac{1}{\pi }\sum_{\sigma }\Im\left[
\partial _{x}a_{\sigma }^{\ast }\partial _{y}a_{\sigma }\right] 
\end{equation}%
which leads\textbf{\ }to 
\begin{eqnarray}
q\left( \mathbf{r}\right)  &=&\frac{1}{4\pi }\widehat{\mathbf{z}}\cdot
\nabla \alpha \times \left[ \nabla \beta -\sin ^{2}\left( \frac{\theta _{R}}{%
2}\right) \nabla \varphi _{R}\right] \sin \alpha   \notag \\
&&+\frac{1}{4\pi }\widehat{\mathbf{z}}\cdot \nabla \alpha \times \left[ \sin
^{2}\left( \frac{\theta _{L}}{2}\right) \nabla \varphi _{L}\right] \sin
\alpha  \\
&&+\frac{1}{4\pi }\left( \widehat{\mathbf{z}}\cdot \nabla \theta _{L}\times
\nabla \varphi _{L}\right) \left[ \sin ^{2}\left( \frac{\alpha }{2}\right)
\sin \theta _{L}\right]   \notag \\
&&+\frac{1}{4\pi }\left( \widehat{\mathbf{z}}\cdot \nabla \theta _{R}\times
\nabla \varphi _{R}\right) \left[ \cos ^{2}\left( \frac{\alpha }{2}\right)
\sin \theta _{R}\right] .  \notag
\end{eqnarray}%
The topological density is related to the charge density\cite{hasebe} so
that a global change in the azimuthal angles has no effect on the charge
density itself.

\section{Conclusion}

In a previous work\cite{cotecp3}, it was shown that a crystal with entangled
spin and pseudospin textures, i.e. a $CP^{3}$ Skyrmion crystal, can be the
Hartree-Fock ground state of the 2DEG in a DQWS when interlayer coherence is
established at small interlayer separation. In the present work, we have
presented an exhaustive study of the collective excitations of such a
crystal by working in the generalized Random-Phase approximation. In the
limiting cases of a pure spin-Skyrmion crystal i.e. at strong tunneling or
at strong bias, we found that the spin-Skyrmion crystal has a gapless phonon
mode and a separate Goldstone mode that arises from a broken $U\left(
1\right) $ symmetry. At zero Zeeman coupling, we demonstrated that the
constituent Skyrmions broke up, and the resulting state is a meron crystal
with \textit{four} gapless modes. In contrast, a pure pseudospin Skyrme
crystal at finite tunneling has only the phonon mode as a Goldstone mode.
For $\Delta _{SAS}\rightarrow 0$, however, the pseudospin Skyrme crystal
evolves into a meron crystal and it then supports an extra gapless ($U(1)$)
mode in addition to the phonon.

For a $CP^{3}$ Skyrmion crystal, we found a $U(1)$ gapless mode (in addition
to the phonon mode)\ in the presence of non-vanishing symmetry-breaking
fields $\Delta _{SAS}$ (tunneling), $\Delta _{Z}$ (Zeeman coupling), and $%
\Delta _{b}$ (electrical bias). This phase mode is explained by the fact
that the energy of the $CP^{3}$ crystal is unchanged if we change the
azimuthal angles $\varphi _{R}$ and $\varphi _{L}$ of \textit{all} the spins
in the right or left layers by the same amount. In addition to this gapless
mode, we found a second mode with a very small gap. We believe that this
mode can be associated with the variation of the relative azimuthal angles $%
\varphi _{R}-\varphi _{L}$ or with the variation of the azimuthal angle $%
\beta $ of the total pseudospin vector or to a combination of these angles.

Having established the low-energy excitations of the $CP^{3}$ Skyrmion
crystal and obtained the corresponding response functions, it is now
possible to compute the $NMR$ relaxation time $T_{1}$ and see how it
compares with the measured experimental values\cite{kumadaprl,spielmanprl}
in order to see if the formation of a $CP^{3}$ crystal plays a role in these
experiments. This, however, is beyond the scope of the present work.

\begin{acknowledgments}
This work was supported by a research grant (for R.C.) from the Natural
Sciences and Engineering Research Council of Canada (NSERC). H.A.F.
acknowledges the support of NSF through Grant No. DMR-0704033. Computer time
was provided by the R\'{e}seau Qu\'{e}b\'{e}cois de Calcul Haute Performance
(RQCHP).
\end{acknowledgments}


\begin{thebibliography}{99}
\bibitem{correction} The factor $1/2$ in Eqs. (\ref{a2}-\ref{a4}) is missing
in the definition of these fields in Paper I\cite{cotecp3}. This is a typo.
The correct definition was used in our calculation.
\end{thebibliography}
\end{document}